\documentclass[eqsecnum,showpacs,pre,twocolumn]{revtex4-1}%
\usepackage{graphicx}
\usepackage{epsfig}
\usepackage{dcolumn}
\usepackage{amsfonts}
\usepackage{amsmath}
\usepackage{amssymb}
\usepackage{bm}%
\usepackage[usenames]{color}
\begin{document}

\newcommand{\brm}[1]{\bm{{\rm #1}}}
\newcommand{\Ochange}[1]{{\color{red}{#1}}}
\newcommand{\Ocomment}[1]{{\color{green}{#1}}}%{{\color{PineGreen}{#1}}}
\newcommand{\Oremove}[1]{{\color{yellow}{#1}}}
\newcommand{\Hcomment}[1]{{\color{blue}{#1}}}%{{\color{ProcessBlue}{#1}}}
\newcommand{\Hchange}[1]{{\color{magenta}{#1}}}%{{\color{BurntOrange}{#1}}}

\title{Linear Polymers in Disordered Media\\
 -- the shortest, the longest and the mean(est) SAW on percolation clusters}
\author{Hans-Karl Janssen}
\affiliation{Institut f\"{u}r Theoretische Physik III,
Heinrich-Heine-Universit\"{a}t, 40225 D\"{u}sseldorf, Germany }
\author{Olaf Stenull}
\affiliation{Department of Physics and Astronomy, University of Pennsylvania,
Philadelphia PA 19104, USA}
\date{\today}

\begin{abstract}
Long linear polymers in strongly disordered media are well described by
self-avoiding walks (SAWs) on percolation clusters and a lot can be learned
about the statistics of these polymers by studying the length-distribution of
SAWs on percolation clusters. This distribution encompasses to distinct
averages, {\em viz}.\ the average over the conformations of the underlying
cluster and the SAW-conformations. For the latter average, there are two basic
options one being static and one being kinetic. It is well known for static
averaging that if the disorder of the underlying medium is weak, this disorder
is redundant in the sense the renormalization group, i.e., differences to the
ordered case appear merely in non-universal quantities. Using dynamical field
theory, we show that the same holds true for kinetic averaging. Our main focus,
however, lies on strong disorder, i.e., the medium being close to the
percolation point, where disorder is relevant. Employing a field theory for the
nonlinear random resistor network in conjunction with a real-world
interpretation of the corresponding Feynman diagrams, we calculate the scaling
exponents for the shortest, the longest and the mean or average SAW to 2-loop
order. In addition, we calculate to 2-loop order the entire family of
multifractal exponents that governs the moments of the the statistical weights
of the elementary constituents (bonds or sites of the underlying fractal
cluster) contributing to the SAWs.  Our RG analysis reveals that kinetic
averaging leads to renormalizability whereas static averaging does not, and
hence, we argue that the latter does not lead to a well-defined scaling limit.
We discuss the possible implications of this finding for experiments and
numerical simulations which have produced wide-spread results for the exponent
of the average SAW. To corroborate our results, we also study the well-known
Meir-Harris model for SAWs on percolation clusters. We demonstrate that the
Meir-Harris model leads back up to 2-loop order to the renormalizable real
world formulation with kinetic averaging if the replica limit is consistently
performed at the first possible instant in the course of the calculation.
\end{abstract}
 \maketitle

%\keywords{Percolation, self-avoiding walks, linear polymers, field theory,
%renormalization group, critical exponents}

\section{Introduction}

Linear polymers in disordered media have been an important topic for
experimental and theoretical study for more than 20 years~\cite{Chak05}. It is
a well known fact that that the universal scaling properties of linear polymers
in strongly disordered media are well described by the statistics of
self-avoiding-walks (SAWs) on percolation clusters. Usually, the term SAW
implicitly refers to a mean or average SAW for which an average is taken over
the number of steps or  intrinsic lengths of all self-avoiding walks with a
specified Eukledian start-to-end distance or, respectively,  over the
start-to-end distances of all self-avoiding walks with a specified intrinsic
length. However, there are other SAWs that are equally interesting. Most
notably, there is the shortest and longest SAW for which the intrinsic length
for given terminal-separation is shortest and longest, respectively. On
critical percolation clusters, all these SAWs are fractals, i.e., their masses
(which are proportional to their intrinsic lengths, of course) as functions of
the start-to-end distance scale with non-integer scaling exponents.
Interestingly, however, it turns out that the mean SAW is more than just a
simple fractal -- it is a multifractal. On critical percolation clusters, the
statistical weights of the elementary constituents (bonds or sites) are
non-trivial, and an entire family of multifractal scaling exponents is required
to characterize the distribution of these weights through its moments.

Although the last 2 decades have brought great advancement in the understanding
of linear polymers in disordered media, there are certain problems that have
caused enduring controversy. For  example it turned out that the famous
Meir-Harris (MH) model \cite{MeHa89} which has long been standing as the only
existing field theoretic model for studying average SAWs on percolation
clusters has trouble with renormalizability \cite{DoMa91,FeBlFoHo04}. Another
example is the puzzling fact that sophisticated numerical simulations by
various groups have produced widespread results for the scaling exponent
$\nu_{\text{SAW}}$ describing the mean length of the average SAW,
see~\cite{Chak05}.

In this paper, we highlight that one has to be careful about the notion of
average SAW if the disorder of the underlying medium is strong. Namely, there
are essentially two qualitatively different ways of averaging over all SAWs
between two connected sites for a given random configuration of a diluted
lattice, one being static and the other being kinetic. It is well known that
the exponent $\nu_{SAW}$ in a non-random medium is the same for static and
kinetic averaging~\cite{MaJaCoSt84,Pe84,KrLy85,Pi85}, and that weak disorder of
the medium in redundant in the sense of the renormalization group (RG) if
static averaging is used \cite{Ha83}. Below, we employ dynamical field theory
to demonstrate that the same holds true for kinetic averaging. To discuss the
effects of strong disorder, we present a renormalizable field theory for SAWs
on percolation clusters based on the random resistor network (RRN). We use this
theory to calculate the scaling exponents of the shortest, the longest and the
average SAW as well as the entire family of multifractal exponents for SAWs on
percolation clusters. This theory demonstrates that kinetic and static
averaging may lead to very different results if the random medium is at the
percolation point. Within the real-world interpretation of Feynman-diagrams,
static averaging leads to non-renormalizability, and hence we argue that the
statistics of linear polymers in disordered media may not have anasymptotic
scaling limit, when static averaging is used. Because the static average has
been used in many simulations, this might explain why the numerical results for
$\nu_{\text{SAW}}$ are so wide spread. In fact, recent simulations by Blavatska
and Janke~\cite{BlaJa2008,BlaJa2009} using kinetic averaging are in excellent
agreement with our theoretical results. To further corroborate our findings and
to shed some light on them from a different angle, we include a discussion of
the MH model in the context of kinetic \emph{vs}.\ static averaging. A brief
account of the work presented here has been given previously in
Ref.~\cite{JaSt2007}.

\section{Observables and averages}
\label{sec:obsAndAves}

The fundamental question addressed in this paper is that of the scaling
behavior the length of a (shortest, longest or mean) SAW on a percolation
cluster when averaged over all cluster conformation. To this end, one can
either consider the scaling of the mean Euclidian distance $R_{L}$ between the
starting point and a random endpoint of an SAW of length $L$,
\begin{equation}
R_{L}\sim L^{\nu_{SAW}}\,,
\end{equation}
or one can study averages over the random length $L(x,y)$ of a SAW (which is
proportional to the number of monomers, the intrinsic length or the mass of the
corresponding polymer) between a pair of sites $(x,y)$ as a function of their
Euklidian distance $\left\vert \mathbf{x}-\mathbf{y}\right\vert$. The latter
approach is more convenient from the standpoint of field theory~\cite{MeHa89},
and we will take it here.

Let $\chi(x,y)$ be the pair connectedness indicator function which is unity if
$x$ and $y$ are connected in the given random configuration $\mathcal{C}$ and
zero otherwise. Let  $Z(w;x,y,\mathfrak{C})$ be the generating function of the
SAWs $\gamma$ with length $L(\gamma)=L$ belonging to the bundle
$\mathcal{B}(x,y;\mathfrak{C})$ of all SAWs starting at $y$ and ending at $x$
both on cluster $\mathfrak{C}$. This generating function can be written as
\begin{align}
Z(w;x,y,\mathfrak{C}) &  =\sum_{L}\mathrm{e}^{-wL}Z_{L}(x,y,\mathfrak{C}%
)\nonumber\\
&  =\sum_{\gamma\in\mathcal{B}(x,y;\mathfrak{C})}p(\gamma)\mathrm{e}%
^{-wL(\gamma)}\,,\label{GenFu}%
\end{align}
where $p(\gamma)$ is a weight function that depends on the averaging procedure
one uses and which, of course, has to satisfy $\sum_{\gamma}p(\gamma)=1$.
Special cases are $p(\gamma)=\delta_{\gamma,\gamma_m}$ if $\gamma_m$ is either
the shortest or the longest SAW belonging to the bundle
$\mathcal{B}(x,y;\mathfrak{C})$. The perhaps most basic averaging procedures
are either static, i.e.,  all the $N_{SAW}^{(\mathcal{B})}$ $\gamma$'s
belonging to  $\mathcal{B}(x,y;\mathfrak{C})$ are weighted equally, $p(\gamma)
= 1/N_{SAW}^{(\mathcal{B})}$, or kinetic, i.e., a given $\gamma$ earns a factor
$1/z$ contributing to $p(\gamma)$ at each ramification where $z-1$ other SAWs
from the bundle $\mathcal{B} (x,y;\mathfrak{C})$ split off and which, in
general, leads to different weights for different $\gamma$'s.Note
that in kinetic averaging, the probabilities satisfy the additivity property
$p(\gamma_1\cup\gamma_2)=p(\gamma_1)+p(\gamma_2)$ when two SAWs $\gamma_1$ and
$\gamma_2$  are identified, e.g., in a coarse-graining procedure. To the
contrary, this is not the case in static averaging because the number
$N_{SAW}^{(\mathcal{B})}$ changes when SAWs are identified, and this fact leads
to trouble when coarse-graining procedures are applied spatial inhomogeneous
fractals like the backbone of a percolation cluster as they are in renormalized
field theory.

In terms of the generating function, the mean length of the SAWs is given by
\begin{equation}
\bigl[L(x,y)\bigr]_{p}=-\frac{\partial}{\partial w}\frac{\bigl[\chi(x,y)\ln
Z(w;x,y,\mathfrak{C})\bigr]_{p}}{\bigl[\chi(x,y)\bigr]_{p}} \, ,
\label{mean-length}%
\end{equation}
where $\bigl[\cdots\bigr]_{p}$ denotes an average over the configurations
$\mathcal{C}$ in which each bond is occupied with probability $p$. At
criticality, one expects scaling behavior of the mean length $M(x,y)$ of long
polymer chains,
\begin{equation}
M(x,y)=\left.  \bigl[L(x,y)\bigr]_{p}\right\vert _{w=w_{c}}\sim\left\vert
\mathbf{x}-\mathbf{y}\right\vert ^{1/\nu_{SAW}}\,.\label{length-scal}%
\end{equation}
For kinetic averaging, in particular, the critical value $w_{c}$ of $w$ is
zero.

It is well known that multifractality can arise when physical processes unfold
on fractals such as critical percolation clusters. Typical examples are
electrical conduction on
RRNs~\cite{RaTaTrBr85,ArReCo85,PaHaLu87,StJa00,StJa01,St00} and random resistor
diode networks~\cite{StJa02,HiStJa02} where the distribution of currents
flowing through the bonds is multifractal, i.e., is characterized by an
infinite set of critical exponents which are not related in a simple linear or
affine fashion. It turns out that the situation is similar for SAWs on
percolation clusters, where the moments
\begin{equation}
L^{(\alpha)}(x,y)=\sum_{b}s_{b}m_{b}^{\alpha} \label{MultiMom}%
\end{equation}
with $s_{b}$ the length of bond $b$ and
\begin{equation}
m_{b}=\sum_{\gamma\in\mathcal{B}(x,y;\mathfrak{C})}\chi_{b}(\gamma
)p(\gamma)\leq1 \label{Gewichte}%
\end{equation}
the statistical weight of bond $b$ with $\chi_{b}(\gamma) =1$ if $b$ belongs to
the SAW $\gamma$ and $\chi_{b}(\gamma) =0$ if it does not, probe distinct
substructures of the underlying percolation cluster and hence the average
\begin{equation}
M^{(\alpha)}(x,y)=\frac{\bigl[\chi(x,y)L^{(\alpha)}(x,y)\bigr]_{p}}%
{\bigl[\chi(x,y)\bigr]_{p}} \label{MDef}
\end{equation}
leads to an infinite family of multifractal exponents $\nu^{(\alpha)}$:
\begin{equation}
M^{(\alpha)}(x,y)\sim\left\vert \mathbf{x}-\mathbf{y}\right\vert
^{1/\nu^{(\alpha)}}\,. \label{MultiFrak}%
\end{equation}
 One has the special cases $\nu^{(0)}=1/D_{bb}$, $\nu
^{(1)}=\nu_{SAW}$, and $\nu^{(\infty)}= 1/D_{red}=\nu$ where $D_{bb}$ is the
fractal dimension of the backbone, $D_{red}$ is the fractal dimension of the
red (simply connecting) bonds, and $\nu$ is the percolation correlation length
exponent.

\section{SAWs in disordered media as a kinetic process -- the effects of weak disorder}
\label{sec:weakDisorder}

In this section we briefly discuss the effects that weak disorder of the
underlying medium has on the statistics of SAWs. For static averaging over SAW
conformations, it well known that this disorder is redundant in the sense of
the RG, i.e., it affects only non-universal quantities (in annealed as well as
in quenched disorder-averages) although a na\"{\i}ve application of the Harris
criterion~\cite{Ha83} apparently signals its relevance. This redundancy can be
shown by introducing non-correlated quenched disorder into the usual
$\varphi^{4}$-field theory of a $n$-component order parameter which, when the
replica trick with $m$-fold replication is used to facilitate averaging over
the quenched disorder, generates the statistics of SAWs with static averaging
via the limits $n\rightarrow0$ and $m\rightarrow0$. Here, we are interested
mainly in kinetic averaging, and hence it is of some interest to demonstrate
that this redundancy also holds for this type of averaging. A field theory for
kinetically generated SAWs in ordered media has been formulated by Peliti more
than twenty years ago \cite{Pe84}. Here, we introduce a dynamical field
theoretic model for kinetically generated SAWs in disordered media, and we
utilize this model to discuss the effects of weak disorder. To set the stage,
we first introduce reaction diffusion processes that model SAWs in disordered
media as a kinetic process and then we jump to the dynamical response
functional for these processes. Additional background and some details of its
derivation using the creation-destruction operator formalism are given in
Appendix~\ref{app:dynamicalFunctionalForSAW}.  It should not go un-noted that
this description of SAWs in disordered media as a kinetic process completely
avoids using the replica trick as well as the zero-component-limit and hence
all the potential difficulties associated with these limits.

On a very basic and intuitive level, kinetic SAWs in disordered media can be
described by a set of simple diffusion and reaction processes.
 As in the problem of Brownian walks, the walkers hop on a
$d$-dimensional cubic lattice from a site $\mathbf{r}_{i}$ to a neighboring one
$\mathbf{r}_{i}+\brm{\delta}$. This is described by a random process
\begin{equation}
A(\mathbf{r})\overset{\lambda}{\rightarrow}A(\mathbf{r}+ \brm{\delta})\,,
\label{Diff}
\end{equation}
where $A(\mathbf{r})$ denotes a random walker at $\mathbf{r}$, and $\lambda$ is
the hopping rate. Self-avoidance is introduced by production of markers
$B(\mathbf{r})$ at $\mathbf{r}$, and destruction of the walkers by interaction
with a marker
\begin{subequations}
\label{React}
\begin{align}
&  A(\mathbf{r})\overset{\alpha}{\rightarrow}A(\mathbf{r})+B(\mathbf{r}
)\,,\label{React1}\\
&  A(\mathbf{r})+B(\mathbf{r})\overset{\beta}{\rightarrow}B(\mathbf{r})\,,
\label{React2}
\end{align}
\end{subequations}
where $\alpha$ and $\beta$ are reaction rates. In addition to these processes,
there is quenched disorder modelled by static traps $C(\mathbf{r})$ with a
random distribution $\rho(\mathbf{r})$, and this disorder acts on the walkers
via a destruction process with rate $\gamma$
\begin{equation}
A(\mathbf{r})+C(\mathbf{r})\overset{\gamma}{\rightarrow}C(\mathbf{r})\,.
\label{Trap}%
\end{equation}
Due to the markers and the traps, walkers living at time $t$ should avoid sites
that they have visited at times $t^{\prime}<t$ as well as sites which are
locations of traps.

These diffusion and reaction processes can be condensed into a field theoretic
functional following the work of Peliti. A series of steps which are sketched
in the Appendix leads to the dynamical response functional
\begin{align}
\mathcal{J}&=\int d^{d}x\bigg\{\lambda\int_{-\infty}^{\infty}dt\,\tilde
{s}\Big[\lambda^{-1}\partial_{t}+\tau-\nabla^{2}\Big]s
\nonumber \\
&+\frac{g}{2}%
\Big[\lambda\int_{-\infty}^{\infty}dt\,\tilde{s}s\Big]^{2}%
\biggr\}\,.\label{SAW-J}%
\end{align}
$s(\brm{x}, t)$ is a field that has its origin in the variable $A$ and encodes
the random position $\brm{x}$ of a walker at time $t$ according to the above
reactions. Its mean value is the probability density finding a walker at these
coordinates.  $\tilde{s}(\brm{x}, t)$ is the corresponding response field that
creates a walker at position $\brm{x}$ at time $t$. The dynamical response
functional is invariant under the duality transformation $s(\brm{x},
t)\leftrightarrow \tilde{s}(\brm{x}, -t)$. $\lambda$ is the usual kinetic
coefficient. The dependence of $\mathcal{J}$ on the disorder is hidden in the
parameter $\tau$ and in the coupling constant $g$. The latter consists
of two parts: a positive part stemming from the self-avoidance and an
additional negative part stemming from the disorder. Note that the fact that
disorder and self-avoidance generate the same type of coupling in the
field-theoretic functional makes the Harris criterion inapplicable for the
problem at hand.

Analyzing the RG flow, it turns out that the fixed point value of $g$ which
corresponds to the limit of asymptotically large SAWs is independent of the
nonuniversal value of $g$ and hence of the disorder fluctuations. Because we
are mainly interested in this limit, we can assume that the sole remaining
disorder-dependence rests in $\tau$. However, we can eliminate $\tau$ from the
dynamical response functional by letting $s(t)\rightarrow s(t)\exp(-\lambda t)$
and $\tilde{s}(t)\rightarrow\tilde {s}(t)\exp(\lambda t)$. This implies that
the disorder-averaged Greens functions -- the probability densities for finding
$N$ walkers at positions $\{\mathbf{x}\}$ at times $\{t\}$ if they are created
at positions $\{\mathbf{y}\}$ at times $\{t^{\prime}\}$, respectively   --
\begin{align}
\overline{G_{N}(\{\mathbf{x},t\},\{\mathbf{y},t^{\prime}\})}=\overline{\left\langle
\prod_{\alpha =1}^N s (\mathbf{x}_\alpha,t_\alpha) \prod_{\beta =1}^N \tilde{s}
(\mathbf{y}_\beta,t^\prime_\beta) \right\rangle^{(c)}} ,
\end{align}
where $\langle  \cdots \rangle^{(c)}$ denotes the cumulants with respect to the
Boltzmann weight $\exp(- \mathcal{J})$ and $\overline{\cdots}$ denotes disorder
averaging, are of the form
\begin{align}
\label{relationDisNondis}
\overline{G_{N}(\{\mathbf{x},t\},\{\mathbf{y},t^{\prime}\})}&=G_{N}%
(\{\mathbf{x},t\},\{\mathbf{y},t^{\prime}\})_{0}
\nonumber \\
&\times \exp\Big(\lambda%
%TCIMACRO{\tsum \limits_{\alpha=1}^{N}}%
%BeginExpansion
{\textstyle\sum\limits_{\alpha=1}^{N}}
%EndExpansion
(t_{\alpha}-t_{\alpha}^{\prime})\Big)\,,
\end{align}
where the index $0$ indicates the Greens functions without any disorder. The
differences $(t_{\alpha}-t_{\alpha}^{\prime})$ are proportional to the lengths
of the corresponding SAWs, and hence, the exponential factors correspond to a
change in the non-universal fugacities in the statistics of the SAWs. Thus,
Eq.~(\ref{relationDisNondis}) reveals that all universal properties remain
unaffected by disorder. This establishes that the universal properties of
kinetically generated and averaged SAWs are independent of weak disorder as
their statically averaged counterparts are~\cite{Ha83}.

\section{Nonlinear random resistor networks}

The RRN is a variant of the usual percolation problem where occupied bonds are
viewed as resistors and open bonds are viewed as insulators. Here, we consider
a nonlinear generalization nRRN of the RRN for which it is well known that the
total resistance between 2 points becomes proportional to the length of the
shortest and longest SAW between those 2 points, respectively, for specific
limits of the nonlinearity. In previous
work~\cite{StJaOe99,JaStOe99,JaSt00,St00}, we have shown that the Feynman
diagrams for RRNs (including their nonlinear generalization) have a real-world
interpretation, i.e., they can be considered as being resistor networks
themselves. Based on this real-world interpretation, we here develop an
intuitive and powerful field-theoretic method to calculate the mean length and
multifractal moments of SAWs on percolation clusters.

To be specific, we consider a $d$-dimensional lattice where each bond is
randomly occupied with probability $p$ by a conductor or empty with probability
$1-p$. At each lattice side $i$ there is a voltage $V_{i}$. The voltage drop at
a bond $(ij)$ between sites $i$ and $j$ obeys a generalized Ohm's
law~\cite{KeSt82/84}
\begin{equation}
V_{j}-V_{i}=\rho_{(ij)}\left\vert I_{i,j}\right\vert ^{r-1}\,I_{i,j} \label{Ohm-V}%
\end{equation}
with a bond resistance $\rho_{(ij)}$. Equivalently, with $s=1/r$ the current
$I_{i,j}$ across the bond is given by
\begin{equation}
I_{i,j}=\sigma_{(ij)}\left\vert V_{j}-V_{i}\right\vert ^{s-1}\,(V_{j}%
-V_{i})\,, \label{Ohm-I}%
\end{equation}
where $\sigma_{(ij)}=\rho_{(ij)}^{-s}$ is the non-linear conductance of the
bond. We restrict ourselves in the following to the case that all occupied
bonds have identical elementary conductances $\sigma=\rho^{-s}$, and the
conductances of the unoccupied bonds are zero. The currents are conserved at
each site and obey Kirchhoff's first law
\begin{equation}
\sum_{j}I_{i,j}+I_{i}=0\,, \label{Kirchhoff_1}%
\end{equation}
where $I_{i}$ is an external current $I_{i}$ flowing into site $i$. If there
are only two ports $x$ and $y$, these currents are given by
$I_{i}=I\bigl(\delta _{i,y}-\delta_{i,x}\bigr)$, where $I$ is the current
resulting from the voltage difference $U=V_{y}-V_{x}$ between the two ports.
The electrical power $P$ dissipated in the network is given by the bilinear
form
\begin{align}
P  &  =\sum_{(ij)}(V_{j}-V_{i})I_{i,j}\nonumber\\
&  =\sum_{(ij)}\sigma_{(ij)}\left\vert V_{j}-V_{i}\right\vert ^{s+1}%
=\sum_{(ij)}\rho_{(ij)}\left\vert I_{i,j}\right\vert ^{r+1}\,,
\label{power_i,v}%
\end{align}
where the two last equalities follow from Ohm's law, Eqs.~(\ref{Ohm-V}) and
(\ref{Ohm-I}). Using Eq.~(\ref{Kirchhoff_1}), one has
\begin{equation}
P=UI=R_{r}(x,y)\left\vert I\right\vert ^{r+1} \label{power}%
\end{equation}
in the two-port case, where $R_{r}(x,y)$ denotes the total resistance of the
network between the two ports. Blumenfeld \emph{et al}.\ \cite{BMHA85/86} have
shown that the special values $r=-\infty$, $-1$, $-0$, $+0$, $1$ and $\infty$
describe physically relevant geometric properties of the diluted lattice. In
particular, it is easily demonstrated that for $r\rightarrow\pm0$ the internal
currents
 at each ramification flow only in the direction of the highest ($r\rightarrow
+0$) or the lowest ($r\rightarrow-0$) voltage gradient (electromotorical force)
thereby mapping out the shortest or the longest SAW between the 2 terminals,
respectively. As a consequence, the resistance $R_{r}(x,y)$ is proportional to
the length of shortest or longest SAW between $x$ and $y$ for $r\rightarrow+0$
or $r\rightarrow-0$, respectively. We will use this fact for calculating the
average length $M(x, y)$ for these SAWs via calculating the total nonlinear
resistance  $R_{r}(x,y)$ averaged subject to the condition, that the two ports
are on the same cluster,
\begin{equation}
M_{r}(x,y)=\bigl[\chi(x,y)R_{r}(x,y)\bigr]_{p}/\bigl[\chi(x,y)\bigr]_{p}\,.
\label{Masse_r}%
\end{equation}
At criticality, the average total nonlinear resistance obeys the power law
\begin{equation}
M_{r}(x,y)\sim\left\vert \mathbf{x}-\mathbf{y}\right\vert ^{1/\nu_{r}}\,.
\label{Def-nu_r}%
\end{equation}
which will allow us to extract the SAW exponents for the shortest and the
longest SAW simply by taking the appropriate limit with respect to $r$.

As far as the average SAW is concerned, the situation is somewhat more subtle.
Obviously, the average length of the average SAW lies in between the average
length of the shortest and the longest SAW, which are, of course, very
different. Since average SAW sits somewhere in this discontinuity at $r=0$, it
is not known how to extract its average length from the nRRN by a limit taking.
To overcome this problem, we developed the idea to study the average SAW by
using the real-world interpretation~\cite{JaStOe99,JaSt00,St00,StJa00/01} of
Feynman diagrams which we will discuss in detail in the following section. For
studying SAWs on percolation clusters, we extend the real-world interpretation
originally developed for studying electrical transport on RRNs in that we put
SAWs on Feynman diagrams. As we will explain in detail below, the task of
calculating the average length of SAWs on percolation clusters then in essence
reduces to calculating the average length of SAWs on Feynman diagrams. For the
average SAW in particular, this approach avoids the aforementioned problems
associated with taking a limit in $r$

The resistance $R_{r}(x,y)$ can be obtained by solving the circuit equations
(\ref{Ohm-I}), (\ref{Ohm-V}), and (\ref{Kirchhoff_1}). The circuit equations
can be viewed as a consequence of the variation principle
\begin{equation}
\frac{\partial}{\partial V_{i}}\Big[\frac{1}{s+1}P(\{V\})-I\bigl(V_{x}
-V_{y})\Big]=0\,, \label{varprinc1}%
\end{equation}
where the power $P$ is expressed purely as a function of the set of all
voltages $\{V\}$, see Eq.~(\ref{power_i,v}). Obviously the network may contain
closed loops. Suppose there is a complete set of independent currents
$\{I^{(l)}\}$ circulating around these loops. Using Kirchhoff's first law
(\ref{Kirchhoff_1}) and Eq.~(\ref{power_i,v}), one can express the electrical
power $P$ entirely as a function of the external current $I$ and the set of
loop currents $\{I^{(l)}\}$. Then, one readily obtains  Kirchhoff's second law
as a consequence of the second variational principle
\begin{equation}
\frac{\partial}{\partial I^{(l)}}P(I,\{I^{(l)}\})=0\,. \label{varprinc2}%
\end{equation}
This equation is used in the following  to determine the loop currents as
linear functions of the external current $I$. For $r>0$, the variation
principle (\ref{varprinc2}) has a unique solution which corresponds, of course,
to the global minimum of the power $P$. Without ambiguity, this solution leads
to $R_{r\rightarrow+0}(x,y)\sim L_{\mathrm{min}}(x,y)$ where $L_{\mathrm{min}}$
denotes the length of the shortest SAW (the chemical length). For $r<0$, the
situation is less straightforward. There exist in general several solutions
which are all local maxima of the power $P$ \cite{BMHA85/86}. Only if one
selects the solution corresponding to the global maximum of $P$, one gets the
length of the longest SAW   via $R_{r\rightarrow-0}(x,y)\sim
L_{\mathrm{max}}(x,y)$. As mentioned above, the average length of the average
SAWs has to lie somewhere in the interval between these two extremal values and
for a correct interpretation of $R_{0}(x,y)$, it seems natural to demand that
it produces the length of the average SAW, $R_{0}(x,y)\sim
L_{\mathrm{SAW}}(x,y)$.

\section{Harris model}

A field theory for the non-linear random network was set up by Harris
\cite{Ha87} in analogy to the field theory of the linear case
\cite{St78,HaLu84/87}. The network is replicated $D$ fold: $V_{i}%
\rightarrow\vec{V}_{i}=(V_{i}^{(1)},\ldots,V_{i}^{(D)})$. One considers the
correlation function $G(x,y;\vec{\lambda})=\langle\Psi_{\vec{\lambda}}%
(x)\Psi_{-\vec{\lambda}}(y)\rangle$ of $\Psi_{\vec{\lambda}}(x)=\exp
(i\vec{\lambda}\cdot\vec{V}_{x})$ with complex currents $i\vec{\lambda}\neq 0$:
\begin{align}
G(x,y;\vec{\lambda})  &  =\Bigl[Z^{-D}\int%
%TCIMACRO{\dprod \limits_{j}}%
%BeginExpansion
{\displaystyle\prod\limits_{j}}
%EndExpansion%
%TCIMACRO{\dprod \limits_{\alpha=1}^{D}}%
%BeginExpansion
{\displaystyle\prod\limits_{\alpha=1}^{D}}
%EndExpansion
dV_{j}^{(\alpha)}\exp\Big(-\frac{1}{s+1}P(\{\vec{V}\})\nonumber\\
&  \qquad\qquad\qquad\qquad+i\vec{\lambda}\cdot(\vec{V}_{x}-\vec{V}%
_{y})\Big)\Bigr]_{p}\,. \label{Def_KorrFu}%
\end{align}
Here $P(\{\vec{V}\})=\sum_{\alpha}P(\{V^{(\alpha)}\})$, and $Z$ is the usual
configuration dependent normalization. In order to be well defined, the
integrations over replicated voltages are augmented with extra weight factors
$\exp(i\omega\vec{V}_{j}^{2})$. Physically, these weights correspond to
grounding all sites via unit capacitors. In this picture, $\omega$ with
$\operatorname{Im}\omega>0$ corresponds to the frequency of the voltages.

Because the electrical power depends only on voltage differences, the
integration over the mean voltage of each independent cluster of connected
conductors leads to current conservation for this cluster
in the limit $\omega\rightarrow0$. It follows that $\vec{\lambda}_{x}%
=\vec{\lambda}_{y}=\vec{\lambda}$ if the ports $x$ and $y$ are connected.
However in the case that $x$ and $y$ are not connected, there arise factors
$\sim\exp(c \, \vec{\lambda}^{2}/i\omega)$, where $c$ is some positive
constant, which go to zero in the limit $i\omega\rightarrow -0$. Here, the
condition $\vec{\lambda}\neq0$ is essential. Then, as the result of this
integration, the pair connectedness indicator function of the two ports
$\chi(x,y)$ is automatically generated along with other factors which go to one
in the limit $D\rightarrow0$. Following the work of Harris \cite{Ha87}, the
integration over the voltage differences can now be done by the saddle-point
approximation if we chose $\lambda ^{(\alpha)}=-iI+\xi^{(\alpha)}$ with
$\sum_{\alpha}\xi^{(\alpha)}=0$ under the conditions $1\ll\rho|I|^{r+1}\ll
D^{-1}$ and $\rho|rI^{r-1}\vec{\xi}^{2}|\ll1$ which indeed can be satisfied
simultaneously in the replica limit $D\rightarrow0$. Note that the saddle-point
equations are identical with the variation principle stated in
Eq.~(\ref{varprinc1}). Thus, the saddle-point is determined by the solution of
the circuit equations (\ref{Ohm-V}), (\ref{Ohm-I}), and (\ref{Kirchhoff_1}),
and according to Eqs.~(\ref{power_i,v}) and (\ref{power}), we obtain
\begin{align}
G(x,y;\vec{\lambda})  &  =\Bigl[\chi(x,y)\exp\Big(\frac{\Lambda_{r}}{r+1}%
R_{r}(x,y)\Big)\Bigr]_{p}\nonumber\\
&  \hspace{-1cm}=\bigl[\chi(x,y)\bigr]_{p}\Big\{1+\frac{\Lambda_{r}}{r+1}%
M_{r}(x,y)+\ldots\Big\}\,, \label{R-KumErz}%
\end{align}
where $\Lambda_{r}=\sum_{\alpha=1}^{D}(-i\lambda^{(\alpha)})^{r+1}$ and where
we dropped a factor that goes to 1 in the limit $D\rightarrow0$. Hence,
$G(x,y;\vec{\lambda})$ is the cumulant generating function for the resistance
$R_{r}(x,y)$ between the connected ports $x$ and $y$. $R_{r}(x,y)$ is
proportional to the elementary resistance $\rho$. Hence, $\lim_{\rho
\rightarrow0}G(x,y;\vec{\lambda})=G(x,y;\vec{\lambda}\rightarrow
0)=\bigl[\chi(x,y)\bigr]_{p}$ is the correlation (connectedness) function of
the percolation problem.

To safely exclude $\vec{\lambda}=0$ from the theory it is useful to resort to a
lattice regularization of the voltage-integrals~\cite{footnote1}. One switches
variables $\vec{V}$ to $\vec{\theta}=\frac{\Delta}{\sqrt{N}}\,\vec{k}$ and
$\vec{\lambda}=\frac{\pi }{\Delta\sqrt{N}}\,\vec{l}$ taking discrete values on
a $D$-dimensional torus, i.e.\ $\vec{k}$ and $\vec{l}$ are chosen to be
$D$-dimensional integers with $-N<k^{(\alpha)},l^{(\alpha)}\leq N$ and
$k^{(\alpha)}=k^{(\alpha )}\operatorname{mod}(2N)$,
$k^{(\alpha)}=k^{(\alpha)}\operatorname{mod}(2N)$. $\Delta$ is a redundant
variable with arbitrary scaling behavior.

After discretization there are $(2N)^{D}-1$ independent state variables per
lattice site, and one introduces the Potts-spins
\begin{equation}
\Phi_{\vec{\theta}}(x)=(2N)^{-D}\sum_{\vec{\lambda}\neq0}\exp\bigl(i\vec
{\lambda}\cdot\vec{\theta}\bigr)\,\Psi_{\vec{\lambda}}(x)=\delta_{\vec{\theta
},\vec{\theta}_{x}}-(2N)^{-D} \label{Potts-Spins}%
\end{equation}
subject to the condition $\sum_{\vec{\theta}}\Phi_{\vec{\theta}}(x)=0$. It is
essential as we already have remarked above that the limit $D\rightarrow0$ is
the first of all involved limits and, in particular, has to be taken before
$N\rightarrow\infty$.

The replication procedure leads to the effective Hamiltonian
\begin{equation}
H_{\mathrm{rep}}=-\ln\Bigl[\exp\Big(-\frac{1}{s+1}P\Big)\Bigr]_{p}\,,
\label{H-rep}%
\end{equation}
which may be expanded in terms of $\Phi_{\vec{\theta}}$ or, equivalently,
$\Psi_{\vec{\lambda}}$
\begin{align}
H_{\mathrm{rep}}  &  =-\sum_{<x,x^{\prime}>}\sum_{\vec{\lambda}\neq0}
K(\vec{\lambda})\,\Psi_{-\vec{\lambda}}(x)\Psi_{\vec{\lambda}}(x^{\prime
})\,\nonumber\\
&  =-\sum_{<x,x^{\prime}>}\sum_{\vec{\theta}}\Phi_{\vec{\theta}}
(x)K(i\partial_{\vec{\theta}})\Phi_{\vec{\theta}}(x^{\prime})\,,
\label{H-rep-exp}%
\end{align}
where $\partial_{\vec{\theta}}$ is the (discrete) gradient in the replica
space. Next the kernel $K(\vec{\lambda})$ is expanded in the limit of large
conductance $\sigma$ (small resistance $\rho)$
\begin{equation}
K(\vec{\lambda})=K_{0}+K_{1}\Lambda_{r}+K_{2}\Lambda_{r}^{2}+\ldots\,,
\label{K-expans}%
\end{equation}
with $K_{n}\sim\rho^{n}$. Therefore, in the limit $\sigma\rightarrow\infty$ we
have $K(\vec{\lambda})\rightarrow K_{0}$, and we recover the $(2N)^{D}$-state
Potts-model which describes percolation in the limit $D\rightarrow0$.

By choosing the redundant variable $\Delta$ appropriately, one can
show that all terms in Eq.~(\ref{K-expans}) with $K_{n>2}$ are irrelevant in
the sense of the renormalization group.  The $K_{n>2}$ merely lead to
corrections to the leading scaling behavior that have been calculated for the
linear RRN, $r \to 1$, in Ref.~\cite{JaSt04}. They do not lead, however, to a
family of crossover exponents as erroneously concluded in
Refs.~\cite{HaLu84/87,Wa89}. Bluntly stated, RRNs are multifractal but they are
not multicritical.

\section{Field theory}
\label{sec:fieldTheory}

To set up a field theoretic Hamiltonian $\mathcal{H}$, one proceeds with the
usual coarse graining step and replace the Potts-spins $\Phi_{\vec{\theta}%
}(x)$ by the order-parameter field $\varphi(\mathbf{x},\vec{\theta})$ defined
on a $d$-dimensional spatial continuum. Constructing all possible relevant
invariants of the symmetry transformations of the model, performing a gradient
expansion and discarding all irrelevant terms, one arrives at the Hamiltonian
\begin{equation}
\mathcal{H}=\int d^{d}x\sum_{\vec{\theta}}\Big[\frac{\tau}{2}\varphi^{2}%
+\frac{1}{2}(\nabla\varphi)^{2}+\frac{w}{2}\varphi\,\bigl(-\vec{\partial
}_{\theta}\bigr)^{r+1}\varphi+\frac{g}{6}\varphi^{3}\Big]\,.
\label{Hamiltonian}%
\end{equation}
Here $\tau$ and $w$ are the strongly relevant critical control parameters and
$\bigl(-\vec{\partial}_{\theta}\bigr)^{r+1}:=\sum_{\alpha}(-\partial
/\partial\theta^{(\alpha)})^{r+1}$. For $w=0$ which corresponds to $\rho=0$,
one has full Potts-symmetry, that is invariance under all permutations of the
symmetric group $S_{(2N)^{D}}$. The $(2N)^{D}$ discrete states of the
order-parameter field $\varphi(\mathbf{x},\vec{\theta})=\sum_{\vec{\lambda
}\neq0}\exp\bigl(i\vec{\lambda}\cdot\vec{\theta}\bigr)\,\psi_{\vec{\lambda}%
}(x)$ transform as the fundamental representation of this permutation group.
This fact is crucial because it is the irreducibility of this representation
that ensures that there is only one invariant second and only one invariant
third order coupling with one unique relevant control parameter $\tau$ and
coupling constant $g$, respectively.

\begin{figure}[ptb]
\includegraphics[width=7cm]{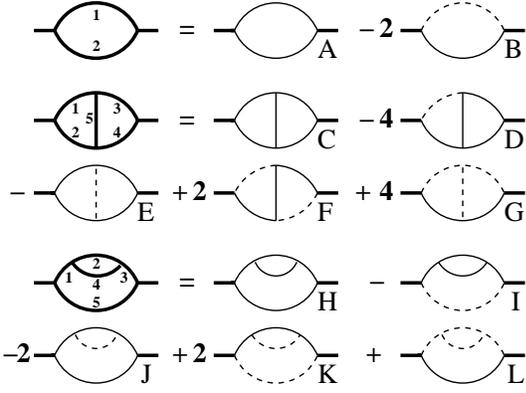}%{decompositionTo2Loop}
\caption{Decomposition of self-energy diagrams to 2-loop order, and enumeration of the lines.}%
\label{Diagrams}%
\end{figure}

Now we set up a diagrammatic expansion. Contributing elements are the vertex
with weight $-g$ and the Gaussian propagator which reads
\begin{equation}
\frac{1-\delta_{\vec{\lambda},0}}{\mathbf{p}^{2}+\tau+w\Lambda_{r}
(\vec{\lambda})}=\frac{1}{\mathbf{p}^{2}+\tau+w\Lambda_{r}(\vec{\lambda}
)}-\frac{\delta_{\vec{\lambda},0}}{\mathbf{p}^{2}+\tau}\ \label{Propag}%
\end{equation}
in Fourier space. This inclusion-exclusion equation shows that the principal
propagator decomposes into a propagator carrying $\vec{\lambda}$ (conducting)
and one not carrying $\vec{\lambda}$ (insulating). This fact allows for a
schematic decomposition of the principal diagrams into sums of RRN-like
conducting diagrams consisting of conducting and insulating propagators, see
Fig.~\ref{Diagrams}. Note that, like the momenta $\mathbf{p}$, the (imaginary)
currents $\vec{\lambda}$ are conserved at each vertex of the conducting
diagrams. For actual calculations, it is more practical to use continuous
rather than discrete replica currents  $\vec{\lambda}$. Once the decomposition
is done, it is save to switch back to integrations over loop replica currents
using
\begin{equation}
\sum_{\vec{\lambda}}\ldots\approx\Big(\frac{\pi}{\Delta\sqrt{N}}\Big)^{D}
\sum_{\vec{\lambda}}\ldots\approx\int%
%TCIMACRO{\dprod }%
%BeginExpansion
{\displaystyle\prod}
%EndExpansion
d\lambda^{(\alpha)}\ldots\label{Sum-Int}%
\end{equation}
where the limit $D\rightarrow0$ is understood.
%%%%%%%%%%%%%%%%%%%%%%%%%%%
%%%%%%%%%%%%%%%%%%%%%%%%%%%

The Feynman diagrams resulting from the decomposition have a simple and
intuitive interpretation~\cite{StJaOe99} which is closely related to the
link-node-blob picture of the back-bone of a percolation cluster. The diagrams
may be viewed as being resistor networks themselves with conducting propagators
corresponding to conductors and insulating propagators to open bonds. In a
Schwinger proper time parametrization where each conducting propagator $i$ is
written as
\begin{equation}
\frac{1}{\mathbf{p}_{i}^{2}+\tau+w\Lambda_{r}(\vec{\lambda}_{i})}=\int
_{0}^{\infty}ds_{i}\,\exp\bigl[-\bigl(\mathbf{p}_{i}^{2}+\tau+w\Lambda
_{r}(\vec{\lambda}_{i})\bigr)\,s_{i}\bigr]\,, \label{Schwinger}%
\end{equation}
the Schwinger parameter $s_{i}$ corresponds to its resistance $\rho_{i}=s_{i}$,
and the replica variables $-i\vec{\lambda}_{i}$ to currents flowing across this
conductor. As mentioned above, this real-world interpretation is closely
related to the link-node-blob picture.The conducting propagators
correspond to the links, the vertices to nodes and the self-energy
diagrams to blobs which themselves can consist of a network of links with
blobs. The backbone corresponds to the full Greens function, i.e.,
the propagator with all possible self-energy insertions summed up by the Dyson
equation. When Fourier transformed back from momentum-space to
configuration-space, the Schwinger parameter (the proper time) of a conducting
propagator is proportional to the intrinsic length of a tortuous link generated
by a diffusional motion, and thus the real-world interpretation naturally
assigns to links their proper length.

The conserved replica currents may be written as
$\vec{\lambda}_{i}=\vec{\lambda}_{i}(\vec{\lambda },\{\vec{\kappa}\})$, where
$\vec{\lambda}$ is an external current applied at the external legs of the
diagram (and subject to the Harris-conditions for application of the saddle
point method) and $\{\vec{\kappa}\}$ denotes the set of independent loop
currents. The replica current dependent part of a diagram can be expressed in
terms of its power $P,$
\begin{equation}
\exp\Big(-w\sum_{i}s_{i}\Lambda_{r}(\vec{\lambda}_{i})\Big)=:\exp
\bigl[-wP(\vec{\lambda},\{\vec{\kappa}\})\bigr]\,. \label{real_world}%
\end{equation}
For the evaluation of the integrals over the independent loop currents we
employ the saddle point method. Note that the saddle point equations constitute
nothing more than the variation principle stated in Eq.~(\ref{varprinc2}).
Thus, solving the saddle point equations is equivalent to determining the total
resistance $R_r(\{s_{i}\})$ of a diagram, and the saddle point evaluation
yields
\begin{equation}
\exp\bigl[-wR_{r}(\{s_{i}\})\Lambda_{r}(\vec{\lambda})\bigr]\,. \label{expR}%
\end{equation}
The Gaussian integration over all internal momenta $\mathbf{p}_{i}$ is textbook
matter. Thereafter, any self-energy diagram, see Fig.~(\ref{Diagrams}), or
rather the mathematical expression standing behind it is of the form
\begin{align}
I(\mathbf{p}^{2},\vec{\lambda})  &  =I_{P}(\mathbf{p}^{2})-wI_{W}%
(\mathbf{p}^{2})\Lambda_{r}(\vec{\lambda})+\ldots\nonumber\\
&  \hspace{-1cm}=\int_{0}^{\infty}%
%TCIMACRO{\dprod \limits_{i}}%
%BeginExpansion
{\displaystyle\prod\limits_{i}}
%EndExpansion
ds_{i}\,\Big[1-wR_{r}(\{s_{i}\})\Lambda_{r}(\vec{\lambda})+\ldots
\Big]\,D(\mathbf{p}^{2},\{s_{i}\})\,, \label{selfen-contr}%
\end{align}
where $D(\mathbf{p}^{2},\{s_{i}\})$ is the $\vec{\lambda}$-independent part of
the integrand which is identical to the integrand of the corresponding diagram
in the standard $\phi^{3}$-theory.

We use dimensional regularization and the renormalization scheme
\begin{align}
\varphi &  \rightarrow\mathring{\varphi}=Z^{1/2}\varphi\,,\qquad
\tau\rightarrow\mathring{\tau}=Z^{-1}Z_{\tau}\tau\,,\nonumber\\
w  &  \rightarrow\mathring{w}=Z^{-1}Z_{w}w\,,\quad g^{2}\rightarrow
\mathring{g}^{2}=G_{\varepsilon}^{-1}Z^{-3}Z_{u}u\mu^{\varepsilon}\,,
\label{RG-Schema}%
\end{align}
where $\varepsilon=6-d$, $\mu$ is an inverse external length scale, and
$G_{\varepsilon}=(4\pi)^{-d/2}\Gamma(1+\varepsilon/2)$ is a factor that
generically emerges in the calculation of Feynman diagrams of a
$\phi^3$-theory. In the limit $D\rightarrow0$, $Z$, $Z_{\tau}$, and $Z_{u}$
are, the well-known percolation renormalizations calculated to three-loop order
by de Alcantara Bonfim, Kirkham, and McKane \cite{AKM81}. It remains to
determine $Z_{w}$ via calculating the part of the self-energy diagrams that is
proportional to $w$, see Eq.~(\ref{selfen-contr}). Note that the
renormalization factors have to fulfill several consistency checks for the
field theory to be renormalizable. Using the fact that the unrenormalized
theory has to be independent of $\mu$, one can set up in a routine fashion a
Gell-Mann--Low RG equation
\begin{align}
&\left[ \mu \frac{\partial }{\partial \mu} + \beta \frac{\partial }{\partial u}
+ \tau \kappa \frac{\partial }{\partial \tau} + w \zeta_r \frac{\partial
}{\partial w} + \frac{N}{2} \gamma \right]
\nonumber \\
&\times G_N \left( \left\{ {\rm{\bf x}} ,w \Lambda_r \left( \vec{\lambda}
\right) \right\} ; \tau, u, \mu \right) = 0
\end{align}
for the connected $N$ point correlation functions $G_N$, where
\begin{mathletters}
\begin{eqnarray}
\label{wilson} \beta \left( u \right) = \mu \frac{\partial u}{\partial \mu}
\bigg|_0 \ , &\quad& \kappa \left( u \right) = \mu \frac{\partial
\ln \tau}{\partial \mu}  \bigg|_0 \ ,\\
\zeta_r \left( u \right) = \mu \frac{\partial \ln w}{\partial \mu}  \bigg|_0 \
, &\quad& \gamma \left( u \right) = \mu \frac{\partial \ln Z}{\partial \mu}
\bigg|_0 \ ,
\end{eqnarray}
\end{mathletters}
are the corresponding Wilson functions. $|_0$ indicates that unrenormalized
quantities are kept fixed while taking the derivatives with respect to $\mu$.
Then, one can use standard methods to solve the RG equation at  the infrared
stable fixed point $u_\ast$, determined by $\beta \left( u_\ast \right) = 0$.
The Gell-Mann--Low function given to 2-loop order by
\begin{align}
\label{WilsonBeta} \beta(u) &  =\mu\partial_{\mu}\left.u\right\vert
_{0}=-\varepsilon
u+\beta(u)\frac{\partial}{\partial u}\ln\bigl(Z^{3}Z_{g}^{2}\bigr)\nonumber\\
&  =\Big(-\varepsilon+\frac{7}{2}u-\frac{671}{72}u^{2}+\ldots \Big)u
\nonumber\\
&=:-\varepsilon u+\beta^{(0)}(u)\,,
\end{align}
leads to
\begin{equation}
\label{fixedPoint}
u_{\ast}=\frac{2}{7}\varepsilon+\frac{671}{3^{2}7^{3}}\varepsilon
^{2}+O(\varepsilon^{3})
\end{equation}
for the fixed point. Augmenting the so-obtained solution with dimensional
analysis, one gets the scaling form
\begin{align}
\label{scaling} &G_N \left( \left\{ {\rm{\bf x}} ,w \Lambda_r \left(
\vec{\lambda} \right) \right\} ; \tau, u, \mu \right)
\nonumber \\
&= \ell^{(d-2+\eta)N/2} G_N \left( \left\{ \ell{\rm{\bf x}} ,
\ell^{-\phi_r/\nu}w \Lambda_r \left( \vec{\lambda} \right) \right\} ;
\ell^{-1/\nu}\tau , u_\ast, \mu \right)  .
\end{align}
$\eta = \gamma_\ast $ and $\nu = \left( 2 - \kappa_\ast \right)^{-1}$ where
$\kappa_\ast = \kappa(u_\ast)$ \emph{etc.}\ are the usual critical exponents
for percolation. $\phi_r = \nu \left( 2 - \zeta_{r\ast} \right)$ is the
resistance exponent. Choosing the flow parameter as $\ell = |{\bf x}-{\bf
x}^\prime |^{-1}$, Taylor-expanding the 2-point function $G_2$ in powers of $w
\Lambda_r ( \vec{\lambda})$  and comparing Eq.~(\ref{R-KumErz}), one finally
obtains
\begin{eqnarray}
\label{MbehavesAs2} M_r ({\rm{\bf x}}, {\rm{\bf x}}^\prime) \sim |{\rm{\bf x}}
- {\rm{\bf x}}^\prime |^{\phi_r / \nu}
\end{eqnarray}
for the scaling behavior of the average total nonlinear resistance.

\section{SAWs on Feynman diagrams}

In previous work, we have applied the real-world interpretation of Feynman
diagrams to calculate the scaling properties of several physically relevant
properties of percolation clusters: their average resistance when the bonds are
linear resistors ($r=1$), the fractal dimensions of the backbone
($r\rightarrow-1$), the minimal (chemical) length ($r\rightarrow+0$), and the
total length of the singly connected (red) bonds ($r\rightarrow\infty$), as
well as the multifractal moments of the current
distribution~\cite{StJaOe99,JaStOe99,JaSt00,StJa00/01,St00}. The key step in
these studies was to  determine the total linear or nonlinear resistance of the
Feynman diagrams (or their multifractal moments in the study of
multifractality) as described above. In all these cases we verified that our
theory was renormalizable. Furthermore, we have checked and verified  our
results were in conformity with results obtained by other methods as far as
those exist.

Now, we extend the real-world interpretation to study SAWs on percolation
clusters. Instead of viewing them as networks on which electrical transport
takes place, we view the Feynman diagrams now as media (or rather the backbones
thereof) on which SAWs take place. In this picture, the conducting propagators
correspond to links that are accessible to the walker and the insulating
propagators are inaccessible. The Schwinger proper time parameter $s_i$ of an
accessible link corresponds to its internal curled length. The essential task
is then to determine the (shortest, longest or average) total lengths
\begin{equation}
L(\{s_{i}\})=\sum_{i}s_{i}m_{i} \, ,
\label{totalDiagramLenght}%
\end{equation}
cf.\ Eq.~(\ref{MultiMom}), of SAWs on the Feynman diagrams. The resulting
mathematical form of the self-energy diagrams, in particular, is that of
Eq.~(\ref{selfen-contr}) with $R_r(\{s_{i}\})$ replaced by $L(\{s_{i}\})$. From
there on, after fixing the weights $m_i$ of the propagators of the diagrams,
the remaining calculation is once again textbook matter. Since $L(\{s_{i}\})$
is a linear form of the Schwinger parameters, this calculation can be
represented diagrammatically through self-energy diagrams with insertions into
the conducting propagators. As indicated above, the length
$L_{\text{min}}(\{s_{i}\})$ and $L_{\text{max}}(\{s_{i}\})$ of the shortest and
longest SAW are proportional to $R_{r\rightarrow+0} (\{s_{i}\})$ and
$R_{r\rightarrow-0} (\{s_{i}\})$, respectively. The length
$L_{\text{ave}}(\{s_{i}\})$ of the average SAW sits in the discontinuity at
$r=0$ and therefore can potentially provide helpful insights for its proper
interpretation.

\subsection{The shortest SAW}

For calculating the average length of the shortest SAW on percolation a
percolation cluster, we determine the total length of the shortest SAWs on
Feynman diagrams. For a given self-energy diagram, that length is
\begin{eqnarray}
L_{\text{min}}(\{s_{i}\}) =\min_{\mbox{\scriptsize SAWs}} \ \sum_{i \in
\mbox{\scriptsize SAWs}} s_i \ ,
\end{eqnarray}
where the minimum is taken over all SAWs on conducting propagators connecting
the external legs of that diagram. Details of the further steps leading from
here to the exponent $\nu_{\mathrm{\min}}$ of the shortest SAW have been given
in previous publications~\cite{JaStOe99,JaSt00,St00}, and we will not repeat
them here. The upshot is that the diagrammatic expansion for the shortest SAW
can be mapped onto that for dynamical percolation, at least to 2-loop order.
This provides for an important consistency check for the real-world
interpretation, and it provides also for a convenient way of calculating
$\nu_{\mathrm{\min}}$ by extracting it from the dynamical exponent $z$ of
dynamical percolation~\cite{Ja85}. The result is
\begin{align}
\nu_{\mathrm{\min}}  &  =\frac{1}{2}+\frac{\varepsilon}{24}+\bigg[\frac
{1231}{2352}+\frac{45}{196}\biggl(\ln2-\frac{9}{10}\ln3\biggr)\bigg]\Big(\frac
{\varepsilon}{6}\Big)^{2}
\nonumber\\
&+\cdots\,. \label{s-Exp-eps}
\end{align}

\subsection{The longest SAW}

In this section we calculate the scaling exponent $\nu_{\max}$ of the longest
SAW on a percolation cluster. As detailed above, the length of the longest SAW
between terminal points $\mathbf{x}$ and $\mathbf{x}^\prime$ is proportional to
the total nonlinear resistance between $\mathbf{x}$ and $\mathbf{x}^\prime$ on
that cluster in the limit $r\to -0$. In the framework of the real-world
interpretation, this means that we can calculate $\nu_{\max}$ to 2-loop order
via determining the total lengths
\begin{eqnarray}
L_{\text{max}}(\{s_{i}\}) =\max_{\mbox{\scriptsize SAWs}} \ \sum_{i \in
\mbox{\scriptsize SAWs}} s_i \ ,
\end{eqnarray}
of the longest SAWs on the different self-energy diagrams depicted in
Fig.~\ref{Diagrams}. Some Details of this calculation are presented in
Appendix~\ref{app:longestSAW}. It results in the renormalization factor
\begin{equation}
Z_{w}=1+\frac{u}{4\varepsilon}+\Big(\frac{15}{32\varepsilon}+\frac{3}%
{128}+\frac{70\ln2-69\ln3}{192}\Big)\frac{u^{2}}{\varepsilon}+O(u^{3})\,,
\end{equation}
This result implies that the Wilson function $\gamma_{w}$ is given by
\begin{align}
\gamma_{w}  & =-\frac{u}{4}-\Big(\frac{3}{64}+\frac{70\ln2-69\ln3}{96}\Big)u^{2}%
+O(u^{3})\,.
\end{align}
Evaluating $\zeta_{w}=\gamma-\gamma_{w}$ at the fixed point~(\ref{fixedPoint})
leads us then readily to our final result
\begin{align}
\nu_{\max}  & = \frac{\nu}{\phi_{-0}}  = \frac{1}{2-\zeta_{w\ast}}\nonumber\\
& =\frac{1}{2}+\frac{\varepsilon}{168}+\Big[\frac{5365}{16464}+\frac{15}%
{28}\Big(\ln2-\frac{69}{70}\ln3\Big)\Big]\Big(\frac{\varepsilon}{6}%
\Big)^{2}
\nonumber \\
&+\cdots
\end{align}
for the inverse fractal dimension of the longest SAW.

\subsection{The average SAW}
\label{subsec:averageSAW}

As we have discussed above, we can apply the static or the kinetic rule to
calculate the average total lenght $L(\{s_{i}\})$ of SAWs on a Feynman diagram.
At one loop order, kinetic and static averaging lead to identical results. At
two loop order, however, the situation changes, because the 2 averaging
procedures lead to different results for diagram $H$ shown in
Fig.~\ref{Diagrams}. Using the numeration of propagator-lines indicated in
Fig.~\ref{Diagrams}, the static rule leads to $L_{H}^{(st)}
(\{s_{i}\})=(s_{2}+s_{4}+s_{5})/3+2(s_{1}+s_{3} )/3$, whereas the kinetic rule
gives $L_{H}^{(kin)} (\{s_{i}\})=(s_{1}+s_{3} +s_{5})/2+(s_{2}+s_{4})/4$.
Having these 2 expressions for the averaged length, it is easy to understand
that the static rule does not lead to a renormalizable theory. It is a basic
fact of renormalization group theory that non-primitive divergencies arising
from sub-integrations of a $1$-loop insertion must be cancelled through the
counter-terms introduced by the renormalization of this $1$-loop insertion.
However, the weights of $L_{H}^{(st)} (\{s_{i}\})$ are not compatible with the
weights arising in the corresponding $1$-loop diagram with counter-term
insertion: crunching the insertion to a point (corresponding to
$s_{2}+s_{4}\rightarrow 0$) leads to $L_{H}^{(st)} (\{s_{i}\}) \to
s_{5}/3+2(s_{1}+s_{3})/3$ which is different from the total length of
the $1$-loop self-energy diagram with a point insertion. For kinetic averaging,
however, the additivity property mentioned in Sec.~\ref{sec:obsAndAves} comes
into play, and crunching the insertion to a point gives $ L_{H}^{(kin)}
(\{s_{i}\}) \to (s_{1}+s_{2}+s_{3})/2$ which is equal to the total length of
the $1$-loop self-energy diagram with a point insertion. Hence, the kinetic
rule produces non-primitive divergencies that are cancelled by the
counter-terms from the $1$-loop renormalization but the static rule does not.
Thus, we have to reject the static rule on grounds of renormalizability, and we
will use the kinetic rule in the following.

The remaining steps in calculating the scaling exponent for the average SAW
proceed as outlined above. For details, we refer to
Appendix~\ref{app:multifractalMoments}, where the formulae for the multifractal
moments reduce to those for the average SAW when we set the multifractal index
$\alpha$ equal to 1. We obtain the renormalization factor
\begin{equation}
Z_{w}=1+\frac{u}{2\varepsilon}+\Big(1-\frac{\varepsilon}{3}\Big)\frac{u^{2}
}{\varepsilon^{2}}+O(u^{3})\, \label{Z_w}%
\end{equation}
for the parameter $w$. Having $Z_w$, it is straightforward to extract the SAW
exponent $\nu_{\mathrm{SAW}}$ as described above. We obtain the
$\varepsilon$-expansion
\begin{equation}
\nu_{\mathrm{SAW}}=\frac{1}{2}+\frac{\varepsilon}{42}+\frac{677}
{2058}\Big(\frac{\varepsilon}{6}\Big)^{2}+\cdots\,. \label{m-Exp-eps}%
\end{equation}

\begin{figure}[ptb]
%\begin{center}
\includegraphics[width=8cm]{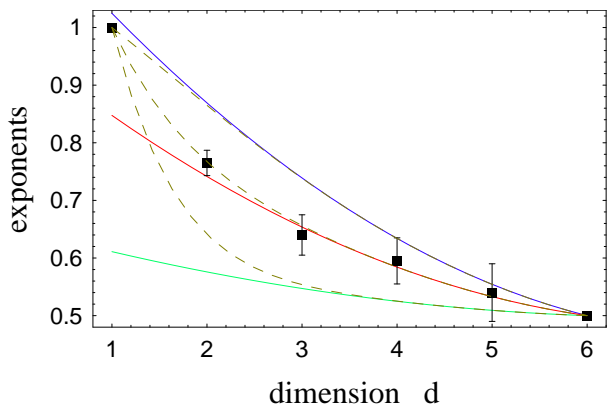}
%\label{Exponenten}(hier)%
%\end{center}
\caption{The $\varepsilon$-expansions of the exponents $\nu_{\mathrm{min}}$
(blue) \cite{Bemerkung}, $\nu_{\mathrm{SAW}}$ (red), and $\nu_{\mathrm{max}}$
(green). The possible extrapolations at low dimensions $d$ are shown by broken
lines. The points denote compiled numerical results for $\nu_{\mathrm{SAW}}$
\cite{Chak05,BlaJa2009}.}%
\label{Exp}%
\end{figure}

For comparison to experimental or numerical data, it is useful to
improve the accuracy of our $\varepsilon$-expansion by implementing rigorously
known features. To this end, we craft  rational approximations for
$\nu_{\mathrm{\min}}$, $\nu_{\mathrm{SAW}}$, and $\nu_{\mathrm{\max}}$ by
adding 8th-order terms in $\varepsilon$ with coefficients chosen such that the
exponents match the rigorously known result $\nu_{...}=1$ in $d=1$.
Table~\ref{tab:exponentValues} compiles numerical values resulting from this
approximation for various dimensions. Figure~\ref{Exp} visualizes our
$\varepsilon$-expansions and rational approximations as functions of $d$. Note
that our rational approximation for $\nu_{\mathrm{SAW}}$ agrees very well with
the available numerical estimates for this exponent which are also shown in
Fig.~\ref{Exp}.
\begin{table}
\begin{tabular}
[c]{c|c|c|c|c|c|c} $d$ & $\quad 1\quad$ & $\quad 2\quad$ & $\quad 3\quad$ & $4$
& $5$ & $6$\\\hline
$\nu_{\min}$ & $1$ & $\; 0.865\;$ & $\; 0.738\;$ & $\; 0.634\;$ & $\; 0.554\;$ & $\; 0.5\;$\\
$\nu_{\mathrm{SAW}}$ & $1$ & $0.767$ & $0.656$ & $0.584$ & $0.533$ & $0.5$\\
$\nu_{\max}$ & $1$ & $0.641$ & $0.554$ & $0.525$ & $0.509$ & $0.5$
\end{tabular}
\caption{\label{tab:exponentValues} Numerical values for various dimensions of
the SAW exponents resulting from rational approximation.}
\end{table}
%%%%%%%%%%%%%%%%%%%%%%%%%%%
%%%%%%%%%%%%%%%%%%%%%%%%%%%

\subsection{Multifractality}
As mentioned above, the fascinating phenomenon of multifractality has been
found in the past in situations where transport processes like electrical
conduction take place on critical percolation clusters. It is reasonable to
expect that multifractality also occurs in the context of SAWs on percolation
clusters, and, indeed, it does~\cite{JaSt2007,BlaJa2008}. The length of the
average SAW that we just computed corresponds to the first of the multifractal
moments defined in Eq.~(\ref{MultiMom}). Now we allow the power $\alpha$ with
which statistical weights of SAWs enter in Eq.~(\ref{MultiMom}) to be arbitrary
positive numbers. Doing so, we can influence the way how SAWs
contribute to the average over their bundle and thereby, loosely speaking, map
out the different fractal substructures of the mean SAW. This approach is
guided by earlier work on RRNs, where multifractality manifests itself in the
moments of the current distribution.

Using the real-world interpretation, we proceed in essentially the same way we
did in Sec.~\ref{subsec:averageSAW} with the only difference that we now
determine all moments of the statistical weights of SAWs on the self-energy
diagrams, i.e., we now keep the $\alpha$ in Eq.~(\ref{MultiMom}) instead of
restricting ourselves to $\alpha=1$. For details of this  calculation, we refer
the reader to Appendix~\ref{app:multifractalMoments}. Here, we would like to
point out, however, that this calculation further underscores the imperative of
kinetic averaging because it leads to a renormalized theory even if all moments
are included whereas static averaging does not. Our calculation produces the
renormalization constants
\begin{align}
\label{renFaktorZalpha}
Z_{\alpha}&=1+\Big(1-\frac{1}{2^{\alpha}}\Big)\frac{u}{\varepsilon}
+\Big[\Big(\frac{9}{\varepsilon}-\frac{47}{12}\Big)-\Big(\frac{11}{\varepsilon}
-\frac{65}{12}\Big)\frac{1}{2^{\alpha}}\nonumber\\
&\qquad\qquad +\Big(\frac{2}{\varepsilon}
-\frac{1}{2}\Big)\frac{1}{4^{\alpha}}\Big]\frac{u^{2}}{4\varepsilon}+O(u^{3})\,,
\end{align}
This result implies the Wilson function
\begin{equation}
\gamma^{(\alpha)} = -\Big(1-\frac{1}{2^{\alpha}}\Big)u
+\Big[\frac{47}{24}-\frac{65}{24\cdot2^{\alpha}}+\frac{1}{4\cdot4^{\alpha}}\Big]u^{2}+O(u^{3})\,,
\end{equation}
Evaluating $\zeta^{(\alpha)}=\gamma-\gamma^{(\alpha)}$ at the fixed
point~(\ref{fixedPoint}), and using $\nu^{(\alpha)} =
1/(2-\zeta^{(\alpha)}_{\ast})$ leads us then readily to our final result
\begin{equation}
\nu^{(\alpha)}=\frac{1}{2}+\Big(\frac{5}{2}-\frac{3}{2^{\alpha}}
\Big)\frac{\varepsilon}{42}+\Big(\frac{589}{21}-\frac{397}{14\cdot2^{\alpha}
}+\frac{9}{4^{\alpha}}\Big)\Big(\frac{\varepsilon}{42}\Big)^{2}+\cdots
\label{MultiFrakExp}%
\end{equation}
for the family of multifractal scaling exponents defined by
Eq.~(\ref{MultiFrak}). As it should, this result for general $\alpha$ reduces
in the special case $\alpha = 1$ to our result for $\nu_{\mathrm{SAW}}$ given
above, and is perfectly consistent with the known results for the backbone and
red bonds dimensions. This can easily be checked by setting $\alpha$ equal to
$0$ and letting $\alpha \to \infty$, respectively.

Note that Blavatska and Janke~\cite{BlaJa2008,BlaJa2009} have devised a
Pad\'{e}-type approximation of our $\varepsilon$-expansion results for the
multifractal exponents which comprise $\nu_{\mathrm{SAW}}$. This approximation
agrees very nicely with their numerical results.

\section{Meir-Harris model}

In this section we discuss in some detail the RG of the Meir-Harris model for
the average SAW on percolation clusters. Our motivation to do so is two-fold.
First, we think that it is of some interest to shed light on the problem at
hand from a different angle, in particular, because we have no rigorous
justification for our real-world interpretation based approach in the form of a
mathematical proof. We will see below, that the MH model when renormalized
properly produces to 2-loop order the same result for the multifractal
exponents as the real-world interpretation and hence provides a strong positive
consistency check for the latter. Second, the RG of the MH model is very
intricate and not properly understood even though the model has existed for
more that 20 years now. A recent 2-loop calculation~\cite{FeBlFoHo04} struggled with
this intricacy and produced incorrect results.

\begin{figure}[ptb]
\centering{\includegraphics[width=2cm]{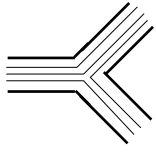}} \caption{Replicons flowing
through a vertex.}%
\label{fig:Repl}%
\end{figure}

It is well known that the statistical properties of SAWs can be calculated from
the $m$-component spin model with $O(m)$-symmetry in the limit $m\rightarrow0$.
To treat dilution, Meir and Harris \cite{MeHa89} start from the $n$-replicated
version of the model. They introduce tensor fields
$\Psi_{k}(\mathbf{x})=\{\Psi_{k;\alpha_{1},\ldots\alpha_{k}}^{\;\;\;\;i_{1}%
,\ldots i_{k}}(\mathbf{x})\}$, $1\leq k\leq n$, conjugate to the product of the
replicated spin-components where the vector-indices $i_{l}$ are running from
$1$ to $m$ and the replica indices $\alpha_{l}=1,\ldots,n$ are arranged such
that $\alpha_{1}<\cdots<\alpha_{k}$. Using the Hubbard-Stratonovich
transformation and passing to the continuum limit, they obtain the effective
Hamiltonian
\begin{equation}
\mathcal{H}=\int d^{d}x\,\Big\{\sum_{k}\Psi_{k}\bigl(r_{k}-\nabla
^{2}\bigr)\Psi_{k}+\frac{g}{6}\Psi^{3}\Big\}\,. \label{MH-Hamiltonian}%
\end{equation}
Here, $\Psi^{3}$ is a symbolic notation for the sum over products of three
$\Psi_{k}$ fields. Only those cubic terms are allowed for which all pairs
$(i,\alpha)$ appear exactly twice. Diagrammatically, this rule can be
represented as shown in Fig.~\ref{fig:Repl} for the $\Psi_{3}\Psi_{3}\Psi_{2}$
coupling. Each of the SAW-representing replicons (thin lines) carries the
indices of the corresponding field $\Psi$. No two pairs of indices entering an
interaction vertex through a given inbound leg are permitted to exit the vertex
through the same outbound leg. Furthermore, the SAW-limit $m\rightarrow0$ for
the $i$-indices implies that diagrams in which some pairs of indices flow in
closed loops produce vanishing contributions. Over all,  any replicons flowing
through an external line into a diagram must flow out off the diagram through
another external line without making any internal loop. Therefore, the basic
task is to count the different distributions of these self-avoiding replicons
under the condition that each line of the diagram bears at least one replicon.
To circumvent the latter condition, it is useful to split each internal fat
line (propagator with replica index $k>0$) into a difference of a conducting
($k\geq0$) and an insulating ($k=0$) line. This step leads to diagrams that can
be drawn in the same way as those for the RRN, see Fig.~\ref{Diagrams}. After
this decomposition, the next step is  draw all possible self-avoiding replicons
on the conducting diagrams where, of course, replicons can flow only through
conducting propagators. Then, one has to sum over all replicon distributions,
i.e.,  all possible arrangements of internal replica indices by given external
ones. Using elementary combinatorics, one finds that this summation for a
diagram with $N$
external legs produces a factor%
\begin{equation}
Z(\{k_{ij}\})=\prod_{(i,j)}\bigl(N_{SAW}(i,j)\bigr)^{k_{ij}}\cdot\prod
_{l}\binom{k_{l}}{\{k_{ll^{\prime}}\}}\,, \label{Distributions}%
\end{equation}
where  $k_{ij}=k_{ji}$ is the number of replicons entering at leg $i$
($i=1,\cdots,N$) and exiting at leg $j$, and $k_{i}=\sum_{j}k_{ij}$ is the
total number of replicons entering at leg $i$ ($i=1,\cdots,N$). $N_{SAW}(i,j)$
is the number of different SAWs which can be drawn
between the pair $(i,j)$ of external legs. $\binom{k_{l}}{\{k_{ll^{\prime}}%
\}}$ is the multinomial coefficient $k_{l}!/(k_{l1}!\cdots k_{lN}!)$. Formula
(\ref{Distributions}) reduces to $N_{SAW}^{k}$ for self-energy diagrams with
$k$ replicons and $N_{SAW}$ different SAWs between the two legs. We
parameterize the temperature-like control parameters $r_{s}$ by%
\begin{align}
r_{s}  &  =\sum_{l=0}^{\infty}\binom{s}{l}v_{l}
 =\tau+s\sum_{l=1}^{\infty}\frac{(-1)^{l-1}}{l}v_{l}+O(s^{2})\,, \label{r-v}%
\end{align}
where $\tau = v_0$. This parametrization facilitates the summation over the
replicon distributions after $r_{s}%
$-insertions in the self-energy diagrams, as well as the limit $k\leq
n\rightarrow0$. The $v_{l}$-part of a $r_{s}$-insertion into the internal line
$p$ of an self-energy diagram, and summation over all distributions of $k$
replicons leads to a factor
\begin{equation}
Z(k,l;p)_{v_{l}}=N_{SAW}^{k}\binom{k}{l}\Big(\frac{N_{SAW}(p)}{N_{SAW}}\Big)^{l}v_{l}\,,
\label{NumSAW}%
\end{equation}
where $N_{SAW}(p)$ is the number of SAWs going through the line $p$. Hence, the
$v_{l}$-insertions \textquotedblleft measure\textquotedblright\ the $l$-th
power of the fraction of all SAWs drawn between the two external legs and going
through the line $p$. In this sense the $v_{l}$ measure multifractal moments of
the diagram using the static rule.
%%%%%%%%%%%%%%%%%%%%%%%%%%%
%%%%%%%%%%%%%%%%%%%%%%%%%%%

However, the MH Hamiltonian (\ref{MH-Hamiltonian}) is not multiplicatively
renormalizable as it stands. The order parameter fields $\Psi_{k}$ belong to
different irreducible tensor-representations of the direct product of the
replica-permutation group $S_{n}$ and the rotation group $SO(m)$ for different
$k$. Hence, the fields $\Psi_{k}$ need $k$-dependent renormalization factors,
and the model is critical at different values \{$r_{k}^{c}\}$ (which are
superficially set to zero in dimensional regularization) of the
temperature-like control parameters. Therefore, the model is highly
multicritical. For an earlier critique concerning this point see Le Doussal and
Machta \cite{DoMa91}. Furthermore, one needs independent coupling constants
$g_{k,l,m}$ for each product of three $\Psi_{k},\Psi_{l},\Psi_{m}$ as opposed
to a single coupling constant $g$ because it is not possible to construct from
the $n$-fold replicated $m$-vector model a higher simple symmetry-group where
the order parameters $\Psi_{k}$ for all $k$ belong to one and the same
irreducible representation unlike in the case of the $n$-fold replicated
$m$-state Potts-model leading to the $m^{n}$-state Potts-model where such a
construction is possible and commonly applied. The latter model, relevant for
the dilute Ising model ($m=2$) and the random resistor network ($m\rightarrow
0$) \cite{HaLu87}, therefore needs only one "scalar" coupling constant $g$ and
a unic (but $m^{n}$-dependent) renormalization factor for all fields, and it is
possible to apply the replica limit at the very end. It is not clear, however,
for the MH model at which stage of its perturbation theory the replica limit
should be taken. There has been hope that if the appropriate stage to take the
replica limit can be identified the renormalizability of the MH model in the
form of a conventional multiplicative renormalization can be
restored~\cite{Ha83b}. In the following, we will embark on a quest to
identify the proper "timing" for the replica limit.

%%%%%%%%%%%%%%%%%%%%%%%%%%%
\begin{figure}[ptb]
\centering{\includegraphics[width=3cm]{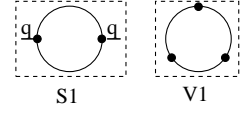}}\caption{$1$-loop
counter-terms.}%
\label{fig:Count-1}%
\end{figure}
%%%%%%%%%%%%%%%%%%%%%%%%%%%
%%%%%%%%%%%%%%%%%%%%%%%%%%%
\begin{figure}[ptb]
\centering{\includegraphics[width=6cm]{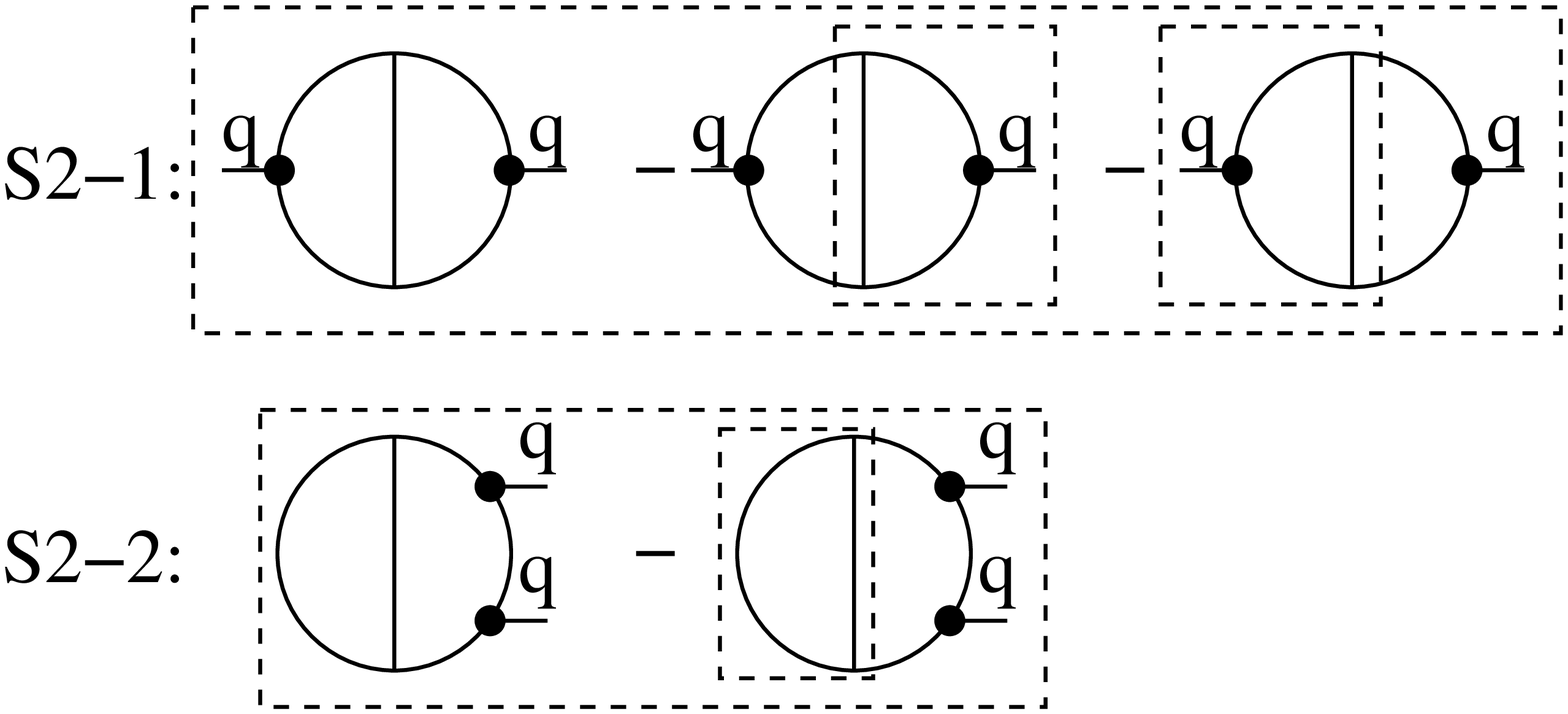}}\caption{$2$-loop
self-energy counter-terms.}%
\label{fig:Count-2-S}%
\end{figure}
%%%%%%%%%%%%%%%%%%%%%%%%%%%
%%%%%%%%%%%%%%%%%%%%%%%%%%%
\begin{figure}[ptb]
\centering{\includegraphics[width=4.5cm]{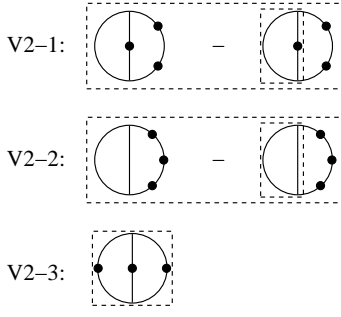}}\caption{$2$-loop
vertex
counter-terms.}%
\label{fig:Count-2-V}%
\end{figure}
%%%%%%%%%%%%%%%%%%%%%%%%%%%

\begin{figure}[ptb]
\centering{\includegraphics[width=5cm]{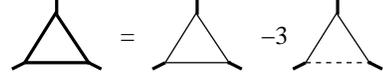}}\caption{Decomposition of the
1-loop vertex diagram.}%
\label{fig:1loopv}%
\end{figure}

One of the most basic facts of RG theory states that non-primitive
divergences arising at a given loop-order in superficially divergent
sub-diagrams must be canceled by the counter-terms resulting from lower
loop-orders. The perhaps most direct route to understand this fact is provided
by the iterative approach to constructing counter-terms invented at the dawn of
RG theory by Bogoliubov, Parasyuk, Hepp, and Zimmermann (BPHZ), see, e.g.,
Ref.~\cite{IZ80}.  We will use a BPHZ-like construction of counter-terms in a
massless $2$-loop calculation using t'Hoofts minimal dimensional
renormalization \cite{tHooft73}. For practical purposes, it is useful to split
up the calculation into a part that determines the counter-terms of frames,
i.e., those parts of Feynman diagrams that stand  only for momentum
integrations, and a part that determines decorations, i.e., symmetry factors,
coupling constants and all other parameters that multiply the frames. There are
7 frame-counter-terms to 2-loop order, see
Figs.~\ref{fig:Count-1}-\ref{fig:Count-2-V}. Our calculation produces
\begin{align}
\text{S1}  &  =-\frac{G_{\varepsilon}\mu^{-\varepsilon}}{3\varepsilon}%
q^{2}\,,\qquad \text{V1}=\frac{G_{\varepsilon}\mu^{-\varepsilon}%
}{\varepsilon}\,,\label{SV1}\\
\text{S2-1}  &  =\frac{G_{\varepsilon}^{2}\mu^{-2\varepsilon}}{3\varepsilon
^{2}}\Big(1-\frac{\varepsilon}{3}\Big)q^{2}\,,
\\
\text{S2-2}&=-\frac
{G_{\varepsilon}^{2}\mu^{-2\varepsilon}}{18\varepsilon^{2}}\Big(1-\frac
{11\varepsilon}{12}\Big)q^{2}\,,\label{S2}\\
\text{V2-1}  &  =-\frac{G_{\varepsilon}^{2}\mu^{-2\varepsilon}}{2\varepsilon
^{2}}\Big(1-\frac{\varepsilon}{4}\Big)\,,\\
\text{V2-2}
&=\frac{G_{\varepsilon}^{2}\mu^{-2\varepsilon}}{6\varepsilon^{2}}\Big(1-\frac
{7\varepsilon}{12}\Big)\,,
\\
\text{V2-3}&=\frac{G_{\varepsilon}^{2}
\mu^{-2\varepsilon}}{2\varepsilon}\,. \label{V2}%
\end{align}
for these counter-terms. To calculate the corresponding decorations using
formulas (\ref{Distributions}) and (\ref{NumSAW}), we determine all possible
self-avoiding replicons on the conducting 1-loop diagrams shown in
Figs.~\ref{Diagrams} and
\ref{fig:1loopv} as described above. We obtain for the $1$-loop self-energy counter-term%
\begin{equation}
\text{Self1}=\frac{g^{2}}{2}\bigl(2^{k}-2\bigr)\cdot\text{S1}=\frac
{u}{6\varepsilon}\bigl(2-2^{k}\bigr)q^{2}\,, \label{Self1}%
\end{equation}
where $u=G_{\varepsilon}\mu^{-\varepsilon}g^{2}$. The $v_{l}$-insertions into
these diagrams result in the counter-terms%
\begin{align}
\text{Ins1}_{l}\cdot\binom{k}{l}v_{l}&=-g^{2}\bigl(2^{k-l}-2\bigr)\binom{k}%
{l}v_{l}\cdot\text{V1}
\nonumber \\
&=\frac{u}{\varepsilon}\bigl(2-2^{k-l}\bigr)\binom{k}%
{l}v_{l}\,, \label{Ins1}%
\end{align}
and the $1$-loop vertex counter-term with $k$, $l$, and $m$ replicons in the
external legs is%
\begin{align}
\text{Vert1}\cdot g&=-g^{3}\bigl(2^{(k+l+m)/2}-3\bigr)N_{k,l,m}\,\cdot
\text{V1}
\nonumber \\
&=\frac{u}{\varepsilon}\bigl(3-2^{(k+l+m)/2}\bigr)N_{k,l,m}g\,,
\label{Vert1}%
\end{align}
where we have used the notation
\begin{equation}
N_{k,l,m}=\frac{k!\,l!\,m!}{\Big[\Big(\frac{k+l-m}{2}\Big)!\,\Big(\frac
{k+m-l}{2}\Big)!\,\Big(\frac{m+l-k}{2}\Big)!\Big]^{2}}\,.
\end{equation}

For the MH model, we use the renormalization scheme
\begin{align}
\Psi&\to \mathring{\Psi}=Z^{1/2}\Psi\,,
\\
v_l&\to\mathring{v}_{l}=Z^{-1}Z_{l}v_{l} \,,
\\
g&\to\mathring{g}=Z^{-3/2}Z_{g}g \,.
\end{align}
To 1-loop order, the above counter terms are related to the renormalization
factors introduced by this scheme via
\begin{align}
Z&=1+\text{Self1}+\ldots\,,
\\
Z_{l}&=1+\text{Ins1}_{l}+\ldots\,,
\\
Z_{g}&=1+\text{Vert1}+\ldots\,,
\end{align}
In the replica limit $n\rightarrow0$ (which implies vanishing external replicon
numbers $k$, $\ldots$) we retrieve the well-known percolation renormalization
factors to 1-loop order. In particular, we retrieve
\begin{equation}
Z_{g}=1+\frac{2u}{\varepsilon}+\ldots\,.
\end{equation}
Note that the renormalization of $\tau$ follows from the identity
$Z_{\tau}=Z_{\infty}$.
%%%%%%%%%%%%%%%%%%%%%%%%%%%
%%%%%%%%%%%%%%%%%%%%%%%%%%%

Now we turn to the $2$-loop counter-terms using the same general approach as
for the 1-loop part of the calculation. From the diagrams shown in the middle
part of Fig.~\ref{Diagrams}, we obtain the counter-term
\begin{align}
\text{Self2-1}&=\frac{g^{4}}{2}\bigl(4^{k}-5\cdot2^{k}+6\bigr)\cdot \text{S2-1}
\nonumber \\
&=\frac{u^{2}}{6\varepsilon^{2}}\Big(1-\frac{\varepsilon}%
{3}\Big)\cdot\bigl(4^{k}-5\cdot2^{k}+6\bigr)\cdot q^{2}\,. \label{Self2-1}%
\end{align}
 The diagrams shown in the lower part of Fig.~\ref{Diagrams} yield
\begin{align}
\text{Self2-2}&=\frac{g^{4}}{2}\bigl(h^{k}-3\cdot2^{k}+3\bigr)\cdot \text{S2-2}
\nonumber \\
&=-\frac{u^{2}}{36\varepsilon^{2}}\Big(1-\frac{11\varepsilon}%
{12}\Big)\cdot\bigl(h^{k}-3\cdot2^{k}+3\bigr)\cdot q^{2}\,. \label{Self2-2}%
\end{align}
Note the placeholder $h$ appearing in this formula. This placeholder reflects
the fact that there are apparently two possible choices for taking the replica
limit, and the result we obtain for diagram H of Fig.~\ref{Diagrams} depends on
this choice. We can let $n\to 0$ in the superficially diverging  $1$-loop
self-energy subdiagram in $H$ either before or after taking the summation over
the replicon distribution. In the first case, $h=2$ whereas $h=3$ in the
second. We will return to the issue of  these choices further below.
%%%%%%%%%%%%%%%%%%%%%%%%%%%
\begin{figure}[ptb]
\centering{\includegraphics[width=4cm]{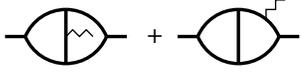}}\caption{Insertions into
the 2-loop self-energy diagram whose decomposition is shown in the middle part
of
Fig.~\ref{Diagrams}.}%
\label{fig:insert2-2}%
\end{figure}
%%%%%%%%%%%%%%%%%%%%%%%%%%%
%%%%%%%%%%%%%%%%%%%%%%%%%%%
\begin{figure}[ptb]
\centering{\includegraphics[width=6cm]{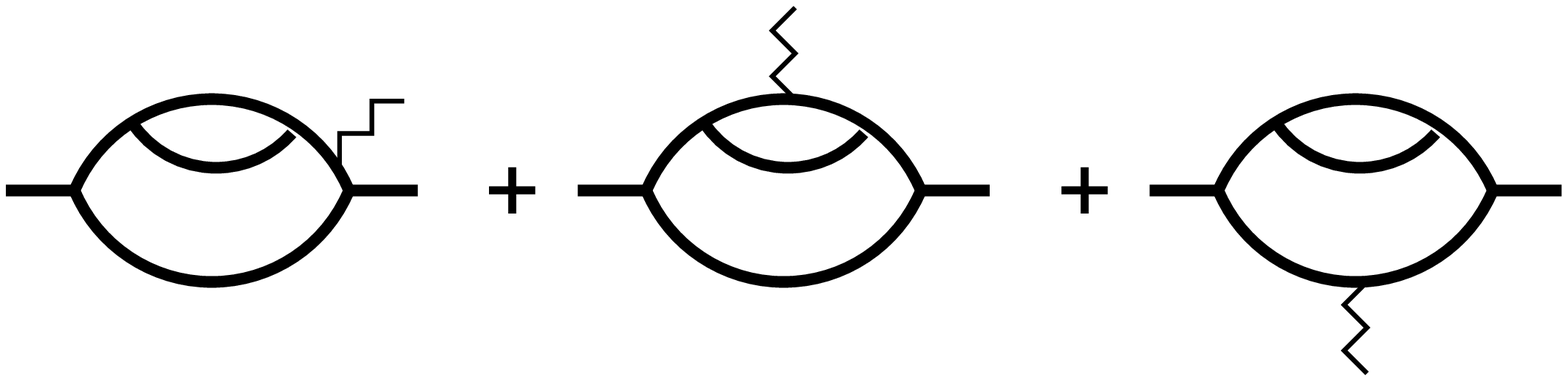}}\caption{Insertions into
the 2-loop self-energy diagram whose decomposition is shown in the lower part
of
Fig.~\ref{Diagrams}.}%
\label{fig:insert2-1}%
\end{figure}
%%%%%%%%%%%%%%%%%%%%%%%%%%%
Next, we consider $v_{l>0}$-insertions into the $2$-loop self-energy diagrams.
From the diagrams shown in the middle part of Fig.~\ref{Diagrams}, we obtain
\begin{align}
&\text{Ins2-1}_{l}\cdot\binom{k}{l}v_{l}
\nonumber \\
&  =-\frac{g^{4}}{2}\bigl(4^{k}%
\cdot(1/2)^{l}-4\cdot2^{k}\cdot(1/2)^{l}+2\bigr)\binom{k}{l}v_{l}%
\cdot\text{V2-3}\nonumber\\
&  -2g^{4}\bigl(4^{k}\cdot(1/2)^{l}-3\cdot2^{k}\cdot(1/2)^{l}-2^{k}%
+3\bigr)\binom{k}{l}v_{l}\cdot\text{V2-1}\nonumber\\
&  =\frac{u^{2}}{\varepsilon^{2}}\Big\{-\bigl(2^{2k-l}-2^{2+k-l}%
+2\bigr)\frac{\varepsilon}{4}\nonumber\\
&  +\bigl(2^{2k-l}-3\cdot2^{k-l}-2^{k}+3\bigr)\Big(1-\frac{\varepsilon}%
{4}\Big)\Big\}\cdot\binom{k}{l}v_{l}\,. \label{Ins2-1}%
\end{align}
Insertions into the $2$-loop self-energy diagrams
shown in the lower part of Fig.~\ref{Diagrams} produce%
\begin{align}
&\text{Ins2-2}_{l}\cdot\binom{k}{l}v_{l}
\nonumber \\
&  =-g^{4}\bigl(h^{k}\cdot
a^{l}-2\cdot2^{k}\cdot(1/2)^{l}-2^{k}+2\bigr)\binom{k}{l}v_{l}\cdot
\text{V2-1}\nonumber\\
&  -g^{4}\bigl(h^{k}\cdot b^{l}-2\cdot2^{k}\cdot(1/2)^{l}+1\bigr)\binom{k}%
{l}v_{l}\cdot\text{V2-2}\nonumber\\
&  -g^{4}\bigl(h^{k}\cdot c^{l}-2\cdot2^{k}\cdot(1/2)^{l}+1\bigr)\binom{k}%
{l}v_{l}\cdot\text{V2-1}\nonumber\\
&  =\frac{u^{2}}{\varepsilon^{2}}\Big\{-\bigl(h^{k}\cdot a^{l}-2^{1+k-l}%
-2^{k}+2\bigr)\frac{1}{6}\Big(1-\frac{7\varepsilon}{12}\Big)\nonumber\\
&  +\bigl(h^{k}\cdot b^{l}-2^{1+k-l}+1\bigr)\frac{1}{2}\Big(1-\frac
{\varepsilon}{4}\Big)\nonumber\\
&  -\bigl(h^{k}\cdot c^{l}-2^{1+k-l}+1\bigr)\Big\}\cdot\binom{k}{l}v_{l}\,,
\label{Ins2-2}%
\end{align}
were $a$, $b$, and $c$  are further $l$-independent placeholders stemming from
the two apparent choices for taking the replica limit as mentioned above. When
we let $n\to 0$ in the $1$-loop superficially diverging self-energy subdiagram
in diagram H before taking the summation over the replicon distribution, we
obtain $a=1/2$, $b=1/4$, $c=1/2$ and $h=2$. Otherwise, we get  $a=2/3$,
$b=1/3$, $c=1/3$ and  $h=3$. We will analyze the correct "timing" of the
replica limit in more detail further below.

To calculate the renormalization factors for the fields to $2$-loop order, we
collect our various diagrammatic results,
\begin{align}
Z  &  =1+\text{Self1}+\text{Self2-1}+\text{Self2-2}+\ldots\,,\\
Z_{l}  &  =1+\text{Ins1}_{l}+\text{Ins2-1}_{l}+\text{Ins2-2}_{l}+\ldots\,,
\end{align}
and take the limit $k\rightarrow0$. In this limit we obtain%
\begin{align}
Z  &  =1+\frac{u}{6\varepsilon}+\Big(11-\frac{37}{12}\varepsilon
\Big)\frac{u^{2}}{36\varepsilon^{2}}+\ldots\,,
\\
Z_{l}  &  =1+\bigl(1-2^{-l}\bigr)\frac{u}{\varepsilon}+\Big[\Big(9-\frac
{47}{12}\varepsilon\Big)-\Big(10-\frac{29}{6}\varepsilon\Big)2^{-l}\nonumber\\
&  -\Big(\frac{2}{3}-\frac{7}{18}\varepsilon\Big)a^{l}+\Big(2-\frac{1}%
{2}\varepsilon\Big)b^{l}-\Big(\frac{1}{3}-\frac{7}{36}\varepsilon
\Big)c^{l}\Big]\frac{u^{2}}{4\varepsilon^{2}}\ldots\,. \label{Z-Faktoren}%
\end{align}
As usual, these renormalization factors, as well as their products have the
form of a Laurent series, $Z=1+\sum _{k=1}^{\infty}Z^{(k)}(u)/\varepsilon^{k}$,
etc.

Our ultimate goal is to determine the inverse multifractal dimensions
\begin{align}
\nu^{(l)}=(2-\kappa_{l\ast})^{-1}
\end{align}
of the MH model. Thus, we need to extract from the above renormalizations the
Wilson function
\begin{equation}
\kappa_{l}=-\beta(u)\frac{\partial}{\partial u}\ln\bigl(Z^{-1}Z_{l}%
\bigr)\,,\label{kappa-l}%
\end{equation}
where $\beta(u)=-\varepsilon u +\beta^{(0)}(u)$ is the Gell-Mann--Low function
given to 2-loop order in Eq.~(\ref{WilsonBeta}).
It follows from Eq.~(\ref{kappa-l}) that%
\begin{align}
\kappa_{l} &  =u\frac{\partial}{\partial u}\bigl(Z^{-1}Z_{l}\bigr)^{(1)}%
-\frac{1}{\varepsilon}\beta^{(0)}(u)\frac{\partial}{\partial u}\bigl(Z^{-1}%
Z_{l}\bigr)^{(1)}\nonumber\\
&  +\frac{1}{2\varepsilon}u\frac{\partial}{\partial u}\Big[2\bigl(Z^{-1}%
Z_{l}\bigr)^{(2)}-\Big(\bigl(Z^{-1}Z_{l}\bigr)^{(1)}\Big)^{2}%
\Big]+O(\varepsilon^{-2})
\end{align}
has to be free of $\varepsilon$-poles. Hence, we obtain the t'Hooft-identity
\cite{tHooft73}
\begin{align}
&u\frac{\partial}{\partial u}\Big[2\bigl(Z^{-1}Z_{l}\bigr)^{(2)}%
-\Big(\bigl(Z^{-1}Z_{l}\bigr)^{(1)}\Big)^{2}\Big]
\nonumber \\
&=\beta^{(0)}(u)\frac
{\partial}{\partial u}\bigl(Z^{-1}Z_{l}\bigr)^{(1)}\,.\label{tH-Identi}%
\end{align}
Inserting our $2$-loop results into this identity, we find the condition%
\begin{equation}
2a^{l}-6b^{l}+c^{l}=3\cdot2^{-l}-6\cdot4^{-l}\,.
\end{equation}
This condition has the unique solution $a=c=2^{-1}$ and $b=4^{-1}$. Thus, to
make the Meir-Harris model renormalizable, one necessarily has to take the
replica limit $n\rightarrow0$ in the superficial divergent subdiagram (SDS)
appearing in diagram H before one sums over the replicon distributions of H
with $v_{l}$-insertion. Remarkably, the renormalization factor $Z_l$ obtained
this way is identical to 2-loop order to the renormalization factor $Z_{w}$
with $ \alpha =l$ resulting from the Harris model in conjunction with the
real-world interpretation provided that kinetic averaging is used.
Consequentially, the same holds true for the multifractal exponents
$\nu^{(\alpha = l)}$ produced by the two approaches. We rate this as a strong
indication for the validity of the real-world interpretation with kinetic
averaging.

\section{Concluding remarks}

In summary, we have shown that weak disorder in the SAW-problem is redundant in
the sense of the RG for kinetic averaging as it is for static averaging.  We
have derived the scaling exponents of SAWs in strongly disordered media by
field-theoretic methods to second order in the dimensional expansion below six
dimensions. We have shown in the real-world interpretation of the corresponding
diagrams that in contrast to a static averaging over the SAWs only kinetic
averaging lead to a renormalizable theory.  The different behavior of these two
averaging procedures under renormalization is expected to have important
physical  consequences for the statistics of polymers in real disordered media.
We argue that a statistics of polymers based on static averaging has no
asymptotic scaling limit. Based on our findings, we do not expect experiments
and numerical simulations using static averaging to produce clear scaling
behavior. In fact, we think that the wide-spreading of simulation results for
the SAW exponent in strongly disordered media is linked to static averaging.

\begin{figure}[ptb]
%\begin{center}
\includegraphics[width=4.0cm]{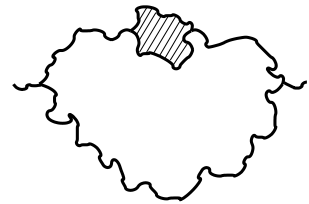}\caption{A blob in a blob of a
percolation cluster in the link-node-blob picture.} \label{nodeLinkBlobCluster}
\end{figure}

Closing, we would like to supplement our firm but conceptually somewhat
involved field-theoretic argument for the imperative of kinetic averaging by a
simple hand-waving argument based on the link-node-blob model of percolation
clusters. In this model, the backbone connecting two terminal points of a
percolation cluster, which is generically very inhomogeneous and asymmetric,
can be envisaged as two nodes linked by a tortuous ribbon that contains blobs.
A blob itself is constructed from at least two links joined at two nodes which
may again contain blobs. Let us for simplicity consider an asymmetric blob as
sketched in Fig.~\ref{nodeLinkBlobCluster} that features two links between two
nodes, one with and the other without a blob.  Note that this cluster resembles
the ominous diagram H of Fig.~\ref{Diagrams}. Assume that the internal blob has
many ramifications of short links in it. Hence, say $N'=N-1\gg 1$ different SAW
configurations are possible on the upper link. With static averaging the upper
link acquires a much larger weight $(N-1)/N$ then the other (lower) one (weight
$1/N$) even if it may be much shorter than the link without the blob. Then, the
statistics of the mean length is dominated by the short upper link with its
many different SAWs induced by the blob. However, the weights change
drastically upon coarse graining. Suppose we have some coarse graining
procedure that culminates in condensing the ``microscopic'' blob into a single
bond. After that, both links have the same weight. However, the lower one,
since it is longer, now dominates the statistics. This demonstrates the
instability of the weights of static averaging under real space renormalization
as the group generated by repeated coarse graining. In contrast, kinetic
averaging does assign the same weight to both links independent of the blob.
Thus, kinetic averaging is stable under coarse graining even in a strongly
asymmetric inhomogeneous disordered medium like the backbone of a percolation
cluster. All in all, the behavior of the links-nodes-blobs cluster of
Fig.~\ref{nodeLinkBlobCluster} under coarse graining resembles in a nut shell
the issues we encountered in our discussion of diagram H of our field theory
and thereby corroborates the imperative of kinetic averaging on an intuitive
level. Note that real-space RG approaches as employed in
Refs.~\cite{MeHa89,FeBlFoHo04} generically use symmetric configurations, and
hence static and kinetic averaging lead to equal results in these approaches as
they do in ordered media like regular lattices which are trivially homogeneous
and symmetric.
%%%%%%%%%%%%%%%%%%%%%%%%%%%
%%%%%%%%%%%%%%%%%%%%%%%%%%%

\begin{acknowledgments}
This work was supported in part (OS) by NSF-DMR-1104707.
\end{acknowledgments}

\appendix

\section{Dynamical response functional for SAWs in random matter}
\label{app:dynamicalFunctionalForSAW}

In this appendix we generalize Peliti's  derivation of a field theory for the
statistics kinetically generated SAWs \cite{Pe84}. An excellent review of the
method can be find in \cite{TaHoVo05}. To have a starting point, let us revisit
the diffusion and reaction processes introduced in Sec.~\ref{sec:weakDisorder}.
First, the rules (\ref{Diff}), (\ref{React}) and (\ref{Trap}) for these
processes are reformulated in terms of a master equation that describes the
time dependence of the probability $P(\{n,m\},t)$ for a given configuration of
site occupation numbers $\{n\}=(\ldots,n_{i},\ldots)$ and
$\{m\}=(\ldots,m_{i},\ldots)$ of the walkers $A$ and the markers $B$,
respectively. Then, the master equation is
transformed in the \textquotedblleft second quantisation\textquotedblright%
\ formalism developed by Doi~\cite{Do76}, Rose~\cite{Ro79}, and Grassberger and
Scheunert~\cite{GrSch80} as follows. The configuration probability is encoded
in the state vector $\left\vert P(t)\right\rangle =\sum_{\{n,m\}}
P(\{n,m\},t)\left\vert \{n,m\}\right\rangle $ in a bosonic Fock space spanned
by the basis $\left\vert \{n,m\}\right\rangle $. These vectors as well as the
stochastic processes in the master equation are expressed through the action of
bosonic creation and annihilation operators $\{a^{+},b^{+}\}$ and $\{a,b\}$,
respectively, which are defined via $a_{i}^{+}\left\vert
\ldots,n_{i},\ldots,\{m\}\right\rangle =\left\vert \ldots,n_{i}+1,\ldots
,\{m\}\right\rangle $ and $a_{i}\left\vert \ldots,n_{i},\ldots
,\{m\}\right\rangle =n_{i}\left\vert \ldots,n_{i}-1,\ldots,\{m\}\right\rangle
,$ etc. $\left\vert 0\right\rangle $ is the vacuum state without walkers and
markers. Subsequently, the master equation can be written in the form%
\begin{equation}
\frac{\partial}{\partial t}\left\vert P(t)\right\rangle =H\left\vert
P(t)\right\rangle \,, \label{SchrGl}%
\end{equation}
with an appropriate non-Hermitean pseudo-Hamilton operator%
\begin{align}
H  &  =\lambda\sum_{<ij>}(a_{j}^{+}-a_{i}^{+})a_{i}
+\sum_{i}\Big\{\alpha(b_{i}^{+}-1)a_{i}^{+}a_{i}
\nonumber \\
&+\beta(1-a_{i}%
^{+})a_{i}b_{i}^{+}b_{i}+\gamma(1-a_{i}^{+})a_{i}\rho_{i}\Big\}\,.
\label{HamOp}%
\end{align}
Here $<ij>$ denotes a pair of neighboring sites. The formal solution of the
master equation (\ref{SchrGl}) reads $\left\vert P(t)\right\rangle
=\exp(tH)\left\vert P(0)\right\rangle $. Suppose we wish to calculate the
$1$-walker probability $P_{1}(i,t)$ to find the walker at site $i$ at time $t$
if he starts from site $0$ at time $0$ not taking into account the resulting
distribution of the markers at $t$. Since a walker can be destroyed but not
spontaneously generated over the
course of time, we have $P_{1}(i,t)=\langle n_{i}\rangle(t)$ where $n_{i}%
=a_{i}^{+}a_{i}$ is the number-operator of walkers at site $i$ and where
$\langle \cdots \rangle$ denotes an average whose precise definition will
become clear shortly. To compute such a statistical average, it is useful to
introduce the projection state $\left\langle \cdot\right\vert =\left\langle
0\right\vert \prod_{i}\exp
(a_{i}+b_{i})$. Using the identities $\left\langle \cdot\right\vert a_{i}%
^{+}=\left\langle \cdot\right\vert $ and $H\left\vert 0\right\rangle =0$ one
easily finds
\begin{equation}
P_{1}(i,t)=\left\langle \cdot\right\vert a_{i}^{+}a_{i}\mathrm{e}^{tH}%
a_{0}^{+}\left\vert 0\right\rangle =\left\langle \cdot\right\vert
a_{i}\mathrm{e}^{tH}(a_{0}^{+}-1)\left\vert 0\right\rangle \,.
\label{1-PartProb}%
\end{equation}

Changing from the occupation-number basis to a Bargmann-Fock space
representation with the coherent states (the eigenstates of the annihilation
operators) as the basis, and following standard procedures
\cite{IZ80,Pe84,TaHoVo05} the expectation value (\ref{1-PartProb}) can be
expressed as a path integral%
\begin{align}
P_{1}(i,t)&=\int\mathcal{D}[\tilde{a},a,\tilde{b},b]a_{i}(t)\tilde{a}%
_{0}(0)\exp\bigl(-S[\{\tilde{a},a,\tilde{b},b\}]\bigr)
\nonumber\\
&=:\langle a_{i}%
(t)\tilde{a}_{0}(0)\rangle\,. \label{PathIntS}%
\end{align}
Here the variables $\{\tilde{a}(t)=a^{+}(t)-1,a(t),\tilde{b}(t)=b^{+}%
(t)-1,b(t)\}$ are classical quantities which correspond to the coherent-state
eigenvalues, and the functional integral (\ref{PathIntS}) is performed subject
to the conditions $\tilde{a}_{i}(\infty)=a_{i}(-\infty)=\tilde{b}_{i}%
(\infty)=b_{i}(-\infty)=0$. The action $S$ results from the Hamiltonian
(\ref{HamOp}) as%
\begin{align}
S  &  =\int_{-\infty}^{\infty}dt\Big\{\sum_{i}\bigl(\tilde{a}_{i}\partial
_{t}a_{i}+\tilde{b}_{i}\partial_{t}b_{i}\bigr)
\nonumber\\
&+\frac{\lambda}{2}\sum
_{<ij>}(\tilde{a}_{i}-\tilde{a}_{j})(a_{i}-a_{j})\nonumber\\
&  +\sum_{i}\Big[-\alpha\tilde{b}_{i}(1+\tilde{a}_{i})a_{i}+\beta\tilde{a}%
_{i}a_{i}(1+\tilde{b}_{i})b_{i}+\gamma\rho_{i}\tilde{a}_{i}a_{i}\Big]\Big\}\,.
\label{Wirk}%
\end{align}
It is easily seen that the coupling induced by the number $1$ in the term
$\alpha\tilde{b}_{i}(1+\tilde{a}_{i})a_{i}$ of the interaction part of
(\ref{Wirk}) does not contribute to the calculation of the expectation values
(Greens functions) $G_{N}(\{i,t\},\{j,t^{\prime}\})=\langle\prod_{\alpha
=1}^{N}a_{i_{\alpha}}(t_{\alpha})\prod_{\beta=1}^{N}\tilde{a}_{j_{\beta}%
}(t_{\beta}^{\prime})\rangle$, especially to $P_{1}(i,t)=G_{1}%
(\{i,t\},\{0,0\})$, by perturbational series. Hence we neglect this destroying
coupling in the following. Furthermore, the variables $b_{i}$ and $\tilde
{b}_{i}$ can be integrated out. \textit{E.g.}, performing the functional
integral $\int\mathcal{D}[b]\exp(-S)$ leads to a factor $\prod_{i,t}%
\delta\bigl(\partial_{t}\tilde{b}_{i}-\beta\tilde{a}_{i}a_{i}(1+\tilde{b}%
_{i})\bigr)$. These $\delta$-conditions can be easily integrated (remember
$\tilde{b}_{i}(\infty)=0$) to%
\begin{equation}
\tilde{b}_{i}(t)=\exp\Big[-\beta\int_{t}^{\infty}dt^{\prime}\,(\tilde{a}%
_{i}a_{i})(t^{\prime})\Big]-1\,.
\end{equation}
The remaining term in $S$ which contains $\tilde{b}_{i}(t)$ yields then%
\begin{align}
-\alpha\int_{-\infty}^{\infty}dt\,\tilde{b}_{i}(t)(\tilde{a}_{i}%
a_{i})(t)&=\alpha A_{i}+\frac{\alpha}{\beta}\tilde{b}_{i}(-\infty)
\nonumber\\
&=\frac {\alpha\beta}{2}A_{i}^{2}+O(A_{i}^{3})\,,
\end{align}
where $A_{i}:=\int_{-\infty}^{\infty}dt\,(\tilde{a}_{i}a_{i})(t)$.

It remains to perform the average of the expectation values $G_{N}$ over the
disorder distribution of the $\rho_{i}$. Note that the normalization factor of
the path integral is defined so that $\int
\mathcal{D}[\tilde{a},a,\tilde{b},b]\exp(-S)=1$, and is purely numeric. Hence,
their is no need for the replica trick. The average can be taken directly over
$\exp(-S)$. We use a Poissonian disorder distribution
\begin{equation}
p(\rho_{i}=k)=\frac{\bar{\rho}^{k}}{k!}\exp(-\bar{\rho})\label{Poiss}%
\end{equation}
of independent traps on each site $i$, characterized by the mean value
$\bar{\rho}$. We get
\begin{align}
&\sum_{k=0}^{\infty}p(k)\exp\Big(-\gamma\int_{-\infty}^{\infty}dt\,k(\tilde
{a}_{i}a_{i})(t)\Big)
\nonumber\\
&  =\exp\Big[\bar{\rho}\Big(\exp(-\gamma A_{i}%
)-1\Big)\Big]
\nonumber\\
&  =\exp\Big[-\bar{\rho}\gamma A_{i}+\frac{\bar{\rho}\gamma^{2}}{2}A_{i}%
^{2}+O(A_{i}^{3})\,.\label{DisAv}%
\end{align}
Using this expression, we arrive at a reduced action for the calculation of the
disorder
averaged Greens functions%
\begin{align}
S_{\mathrm{red}} &  =\int_{-\infty}^{\infty}dt\Big\{\sum_{i}\tilde{a}%
_{i}\partial_{t}a_{i}+\frac{\lambda}{2}\sum_{<ij>}(\tilde{a}_{i}-\tilde{a}%
_{j})(a_{i}-a_{j})\nonumber\\
&  +\sum_{i}\Big[\bar{\rho}\gamma A_{i}+\frac{\alpha\beta-\bar{\rho}\gamma
^{2}}{2}A_{i}^{2}+O(A_{i}^{3})\Big]\Big\}\,.\label{Sred}%
\end{align}
As long as $\alpha\beta-\bar{\rho}\gamma^{2}>0$ the third order terms
$O(A_{i}^{3})$ becomes irrelevant in the RG sense.

To obtain a proper field theoretic functional, it remains to transcribe the
formulation from the lattice to the spatial continuum.  Performing  a naive
continuum limit of $S_{\mathrm{red}}$, several rescalings and a renaming of the
variables and parameters, we finally arrive at  the dynamical response
functional~(\ref{SAW-J}).

\begin{widetext}

\section{The renormalization of the longest SAW}
\label{app:longestSAW}

Here, we present some details of our diagrammatic calculation for the longest
SAW. The general formula of the diagrammatic contributions related to the
various SAWs we consider in this paper is
\begin{equation}
I(\mathbf{q}^{2},\tau,w\lambda)=\int_{0}^{\infty}\prod_{i}ds_{i}%
\,D(\mathbf{q}^{2},\tau,\{s_{i}\}\exp\bigl[-i\lambda w L (\{s_{i}%
\})\bigr]\,, \label{allgemein}%
\end{equation}
where $\{s_{i}\}$ is the set of Schwinger-parameters of a given diagram.
$\lambda$ is defined through
$\Lambda_{0}(\vec{\lambda})=-i\sum_{\alpha=1}^{D}\lambda :=-i\lambda$.
$L(\{s_{i}\})$ is a placeholder for the shortest length
$L_{\text{min}}(\{s_{i}\})$, the longest length $L_{\text{max}}(\{s_{i}\})$ or
the average length $L_{av}(\{s_{i}\})$, respectively, and $w$ is a shorthand
for the corresponding limit of $w_r$. Here, we will focus on
$L_{\text{max}}(\{s_{i}\})$. The average SAW and the corresponding multifractal
moments are treated in the following appendix.

First, let us focus on the 1-loop part of the calculation. For simplicity, we
set $\mathbf{q}=0$ in the following because we are not interested in
reproducing the well-known field-renormalization $Z$. Moreover, we will neglect
all contributions to diagrams proportional to $\tau$ because we are not
interested in  reproducing the well-known $Z_\tau$. At 1-loop order, there are
only 2 conducting diagrams, namely diagrams A and B of Fig.~(\ref{Diagrams}).
These diagrams give
\begin{align}
S_{0}  &  =\text{A - 2B}=\frac{g^{2}}{2}\int_{0}^{\infty}ds_{1}%
ds_{2}\frac{\exp\bigl[-(s_{1}+s_{2})\tau\bigr]}{\bigl[4\pi(s_{1}%
+s_{2})\bigr]^{d/2}} \Big[\exp\bigl[-i\lambda w\max(s_{1},s_{2})\bigr]-2\exp
\bigl[-i\lambda ws_{1}\bigr]\Big]\nonumber\\
&  =-\frac{g^{2}}{2}\int_{0}^{\infty}ds_{1}ds_{2}\frac{\exp\bigl[-(s_{1}%
+s_{2})\tau\bigr]}{\bigl[4\pi(s_{1}+s_{2})\bigr]^{d/2}}\Big[1-2i\lambda
ws_{1}\theta(s_{2}-s_{1})+O(w^{2})\Big]
=\frac{G_{\varepsilon}g^{2}}{\varepsilon}\tau^{-\varepsilon/2} \,
\frac{i\lambda w}{4}\,,
\end{align}
where $w = w_{-0}$ and where $\lambda$ is defined through
$\Lambda_{-0}(\vec{\lambda})=-i\sum_{\alpha=1}^{D}\lambda :=-i\lambda$. Using
the renormalization scheme (\ref{RG-Schema}), we get the $w$-part of the
renormalized selfenergy
\begin{align}
\left.  \Gamma_{2}\right\vert _{w}^{(1l)} &  =Z\left.  \mathring{\Gamma}
_{2}\right\vert _{w}^{(1l)}=i\lambda wZ_{w}\Big[1-Z^{-3+\varepsilon/2}Z_{\tau
}^{-\varepsilon/2}Z_{u}\frac{u}{4\varepsilon}\Big(\frac{\mu^{2}}{\tau
}\Big)^{\varepsilon/2}\Big]\nonumber\\
&  =i\lambda wZ_{w}\Big\{1-\Big[1+\Big(\frac{7}{2\varepsilon}-\frac{5}
{12}\Big)u\Big]\frac{u}{4\varepsilon}\Big(\frac{\mu^{2}}{\tau}
\Big)^{\varepsilon/2}+O(u^{3})\Big\}\,.
\end{align}
It follows the renormalization factor $Z_{w}$ to $1$-loop order:
\begin{equation}
Z_{w}=1+\frac{u}{4\varepsilon}+O(u^{2})\,.
\end{equation}

Now, we turn to $2$-loop order. First, we consider the diagrams C, D, E, F, G.
These diagrams lead to the integral
\begin{align}
S_{1} &  =\text{C - 4D - E + 2F +4G}
=\frac{g^{4}}{(4\pi)^{d}}\int\prod_{i=1}^{5}ds_{i}\,\frac{\Theta
_{1}(\{s_{i}\})\exp\bigl(-\tau\sum_{i}s_{i}\bigr)}{\bigl[(s_{1}+s_{2}%
)(s_{3}+s_{4})+s_{5}(s_{1}+s_{2}+s_{3}+s_{4})\bigr]^{d/2}}\,,
\end{align}
where we have used some invariance under permutations of indices to reduce the
number
of terms and where we have defined%
\begin{align}
\Theta_{1}(\{s_{i}\}) &  =\exp\bigl[-i\lambda w(s_{1}+s_{3})\bigr]\Big\{\theta
(s_{1}-s_{2}-s_{5})\theta(s_{3}-s_{4}-s_{5}) -2\theta(s_{3}-s_{4}-s_{5})-\theta(s_{1}+s_{3}%
-s_{2}-s_{4})+2\Big\}
\nonumber\\
+ &  \exp\bigl[-i\lambda w(s_{1}+s_{4}+s_{5})\bigr]\Big\{\theta(s_{1}%
+s_{5}-s_{2})\theta(s_{4}+s_{5}-s_{3})\theta(s_{1}+s_{4}-s_{2}-s_{3}%
)-2\theta(s_{4}+s_{5}-s_{3})+1\Big\}\,.
\end{align}
To simplify the integrations, we introduce new variables,
\begin{align}
\label{varChange1}
s_{1} &  =xt_{1}\,,\qquad s_{2}=(1-x)t_{1}\,,\nonumber\\
s_{3} &  =yt_{2}\,,\qquad s_{4}=(1-y)t_{2}\,,\qquad s_{5}=t_{3}\,.
\end{align}
The integrations over $x$ and $y$ are cumbersome but manageable and produce
after expansion to linear order in $w$
\begin{align}
S_{1} &  =\frac{g^{4}}{(4\pi)^{d}}\int\prod_{i=1}^{3}dt_{i}%
\,\frac{\exp\bigl(-\tau(t_{1}+t_{2}+t_{3})\bigr)}{\bigl[t_{1}t_{2}+t_{2}%
t_{3}+t_{3}t_{1}\bigr]^{d/2}}\Big\{t_{1}t_{2}+\frac{i\lambda w}{12}%
\bigl(t_{1}t_{2}t_{3}-9t_{1}^{2}t_{2}\bigr)\\
&  \qquad+\frac{i\lambda w}{12}\theta(t_{1}-t_{2})\theta(t_{2}-t_{3}%
)\bigl[3t_{1}^{2}(t_{2}+t_{3})+3t_{2}t_{3}^{2}+t_{2}^{3}+2t_{3}^{3}%
\bigr]\Big\}\,.
\end{align}
In the same manner we find for the second group of diagrams, H, I, J, K, L,
\begin{align}
S_{2} &  =\text{H - I - 2J + 2K +L}
=\frac{g^{4}}{(4\pi)^{d}}\int\prod_{i=1}^{5}ds_{i}\,\frac{\Theta
_{2}(\{s_{i}\})\exp\bigl(-\tau\sum_{i}s_{i}\bigr)}{\bigl[(s_{1}+s_{3}%
+s_{5})(s_{2}+s_{4})+s_{2}s_{4}\bigr]^{d/2}}\,,
\end{align}
with%
\begin{align}
\Theta_{2}(\{s_{i}\}) &  =\exp\bigl[-i\lambda ws_{5}\bigr]\Big\{\frac{1}%
{2}\theta(s_{5}-s_{1}-s_{2}-s_{3})\theta(s_{5}-s_{1}-s_{4}-s_{3})
-\theta(s_{5}-s_{1}-s_{2}-s_{3})+\frac{1}{2}%
\Big\}\nonumber\\
&  +\exp\bigl[-i\lambda w(s_{1}+s_{2}+s_{3})\bigr]\Big\{\theta(s_{1}%
+s_{2}+s_{3}-s_{5})\theta(s_{2}-s_{4})
-\theta(s_{1}+s_{2}+s_{3}-s_{5})-\theta(s_{2}%
-s_{4})+1\Big\}\,.
\end{align}
Here, we chose new integration variables%
\begin{align}
\label{varChange2}
s_{1} &  =(1-x)yt_{1}\,,\quad s_{3}=(1-x)(1-y)t_{1}\,,\nonumber\\
s_{2} &  =t_{2}\,,\qquad s_{4}=t_{3}\,,\qquad s_{5}=xt_{1}\,,
\end{align}
and the integration over $x$ and $y$ yields%
\begin{align}
S_{2} &  =\frac{g^{4}}{(4\pi)^{d}}\int\prod_{i=1}^{3}dt_{i}%
\,\frac{\exp\bigl(-\tau(t_{1}+t_{2}+t_{3})\bigr)}{\bigl[t_{1}t_{2}+t_{2}%
t_{3}+t_{3}t_{1}\bigr]^{d/2}}\Big\{\frac{1}{4}t_{1}^{2}-\frac{i\lambda w}%
{12}t_{1}^{3}
+\frac{i\lambda w}{24}\theta(t_{1}-t_{2})\theta(t_{2}-t_{3}%
)\bigl[(t_{1}-t_{3})^{3}+(t_{2}-t_{3})^{3}\Big\}\,.
\end{align}
Next we turn to the integration over the $t$-variables. In part, these
integrations can be done in an efficient and elegant manner by taking
derivatives of the parameter integral
\begin{align}
M(a,b,c)&=\frac{1}{(4\pi)^{d}}\int\prod_{i=1}^{3}dt_{i}\,\frac{\exp
\bigl(-at_{1}-bt_{2}-ct_{3})\bigr)}{\bigl[t_{1}t_{2}+t_{2}t_{3}+t_{3}%
t_{1}\bigr]^{d/2}}
\nonumber\\
&=
 \frac{G_\epsilon^2}{6\epsilon} \Bigg\{ \left( \frac{1}{\epsilon} + \frac{25}{12}
\right) \left( a^{3-\epsilon} + b^{3-\epsilon} + c^{3-\epsilon} \right)
 - \left( \frac{3}{\epsilon} + \frac{21}{4} \right) \left[ a^{2-\epsilon} \left(
b + c \right) + b^{2-\epsilon} \left( a + c \right) + c^{2-\epsilon} \left( a +
b \right) \right] - 3abc \Bigg\}
\end{align}
introduced by Breuer and Janssen~\cite{BrJa81}. The remaining parts can be
tackled in the same spirit by introducing a second parameter integral:
\begin{equation}
N(a,b,c)=\frac{1}{(4\pi)^{d}}\int\prod_{i=1}^{3}dt_{i}\,\frac{\exp
\bigl(-at_{1}-bt_{2}-ct_{3})\bigr)}{\bigl[t_{1}t_{2}+t_{2}t_{3}+t_{3}%
t_{1}\bigr]^{d/2}}\theta(t_{1}-t_{2})\theta(t_{2}-t_{3})\,.
\end{equation}
In $\varepsilon$-expansion, we obtain
\begin{align}
N(a,b,c) &  =\frac{G_{\varepsilon}^{2}}{6\varepsilon}\bigg\{\frac
{1}{\varepsilon}\Big(\frac{1}{2}a-\frac{9}{4}b-\frac{3}{4}%
c\Big)a^{2-\varepsilon}
  +\Big(\frac{35}{24}-\frac{1}{4}\ln3\Big)a^{3}+\Big(-\frac{1}{3}+\frac{1}%
{2}\ln3-\frac{1}{2}\ln2\Big)b^{3}
 +\Big(-\frac{1}{12}-\frac{1}{4}\ln3+\frac{1}{2}\ln2\Big)c^{3}%
\nonumber\\
& +\Big(-\frac{61}{16}+\frac{9}{8}\ln3-\frac{3}{4}\ln2\Big)a^{2}b +
\Big(-\frac{1}{4}-\frac{3}{2}\ln3+\frac{3}{2}\ln2\Big)ab^{2}%
+\Big(-\frac{31}{16}+\frac{3}{8}\ln3+\frac{3}{4}\ln2\Big)a^{2}c
\nonumber\\
&
  +\Big(\frac{1}{2}+\frac{3}{8}\ln3-\frac{3}{2}\ln2\Big)ac^{2}+\Big(\frac
{1}{2}-\frac{3}{2}\ln3+\frac{3}{2}\ln2\Big)b^{2}c
  +\Big(-\frac{1}{4}+\frac{9}{8}\ln3-\frac{3}{2}\ln2\Big)bc^{2}-\frac{1}%
{2}abc\biggr\}\,.
\end{align}
Via differentiating the parameter integrals with respect to their parameters,
we get the $w$-parts
\begin{align}
\left. S_{1}\right\vert _{w}=-\frac{G_{\varepsilon}^{2}g^{4}\tau^{-\varepsilon}%
}{12\varepsilon} \, i\lambda w\bigg[\frac{6}{\varepsilon}+\Big(-\frac{1}{4}-\ln2+\frac{21}%
{8}\ln3\Big)\bigg]\,,
\end{align}
and%
\begin{align}
\left. S_{2}\right\vert _{w}=-\frac{G_{\varepsilon}^{2}g^{4}\tau^{-\varepsilon}%
}{12\varepsilon} \, i\lambda w\bigg[-\frac{3}{8\varepsilon}+\Big(-\frac{41}{32}-\frac{27}{8}%
\ln2+\frac{27}{16}\ln3\Big)\bigg] \,.
\end{align}
Collecting the 1- and 2-loop contributions, we find
\begin{align}
\left.  \Gamma_{2}\right\vert _{w}^{(2l)}=i\lambda
w\bigg\{Z_{w}-\Big[\frac{1}{4}+\Big(\frac{15}{16\varepsilon
}-\frac{5}{48}\Big)u\Big]\frac{u}{\varepsilon}\Big(\frac{\mu^{2}}{\tau
}\Big)^{\varepsilon/2}
+\Big[\frac{15}{32\varepsilon}+\Big(-\frac{49}{384}-\frac{35}{96}%
\ln2+\frac{69}{192}\ln3\Big)u\Big]\frac{u^{2}}{\varepsilon}\Big(\frac{\mu^{2}%
}{\tau}\Big)^{\varepsilon}\biggr\}
\end{align}
for the $w$-part of the renormalized self-energy to order $u^{2}$. This form
makes evident that non-primitiv divergencies drop out and the
$\varepsilon$-poles are cancelled by choosing
\begin{align}
Z_{w}  & =1+\frac{u}{4\varepsilon}+\Big(\frac{15}{32\varepsilon}+\frac{3}%
{128}+\frac{70\ln2-69\ln3}{192}\Big)\frac{u^{2}}{\varepsilon}+O(u^{3})\,,
\end{align}

\section{Renormalization of the multifractal moments}
\label{app:multifractalMoments}

Now, we present some details of our diagrammatic calculation for the
multifractal moments. Our calculation here is based on Eq.~(\ref{allgemein})
with $L(\{s_{i}\})$ specified to
\begin{equation}
L_{av}(\{s_{i}\})=\sum_{\gamma\in\mathcal{B}}p(\gamma)L_{\gamma}%
(\{s_{i}\})=\sum_{i}m_{i}s_{i}\,, \label{AvSAW}%
\end{equation}
where $L_{\gamma}(\{s_{i}\})=\sum_{i\in\gamma}s_{i}$ is the length of SAW
$\gamma$ of the bundle $\mathcal{B}$ of all SAWs on the conducting part of the
diagram, $p(\gamma)$ its probability, $m_{i}$ the statistical weight of line
$i$ of the diagram. More generally, we are interested in all averaged moments
of the weights, and hence we consider
\begin{equation}
L_{av}^{(\alpha)}(\{s_{i}\})=\sum_{i}m_{i}^{\alpha}s_{i}\,, \label{AvMom}%
\end{equation}
with $\alpha$ left general. For the $1$- and $2$-loop diagrams $A$ to $L$ shown
in Fig.~(1) the probabilities of the SAWs according to the kinetic rule are
\begin{align}
p(\{\gamma\})^{(A)}  &  =\{1/2,1/2\}\,,\qquad p(\{\gamma\})^{(B)}%
=\{1\}\,,\nonumber\\
p(\{\gamma\})^{(C)}  &  =\{1/4,1/4,1/4,1/4\}\,,\quad p(\{\gamma\})^{(D)}%
=p(\{\gamma\})^{(E)}=\{1/2,1/2\}\,,\nonumber\\
p(\{\gamma\})^{(F)}  &  =p(\{\gamma\})^{(G)}=\{1\}\,,\nonumber\\
p(\{\gamma\})^{(H)}  &  =\{1/2,1/4,1/4\}\,,\quad p(\{\gamma\})^{(I)}%
=p(\{\gamma\})^{(J)}=\{1/2,1/2\}\,,\nonumber\\
p(\{\gamma\})^{(K)}  &  =p(\{\gamma\})^{(L)}=\{1\}\,. \label{Dia-SAW-prob}%
\end{align}
Note that only the three SAWs on diagram $H$ the kinetic rule lead to different
probabilities than the static rule which yields $p(\{\gamma
\})_{stat}^{(H)}=\{1/3,1/3,1/3\}$. This fact is discussed in detail in the main
text. The statistical weights of the lines  of the different diagrams follow
from Eq.~(\ref{AvSAW}) as
\begin{align}
&  m_{1}^{(A)}=m_{2}^{(A)}=\frac{1}{2}\,,\qquad m_{2}^{(B)}=1\,,\nonumber\\
&  m_{i=1,2,3,4,5}^{(C)}=m_{3,4,5}^{(D)}=m_{1,2,3,4}^{(E)}=\frac{1}%
{2}\,,\nonumber\\
&  m_{2,3,5}^{(F)}=m_{2,3,5}^{(G)}=1\,,\nonumber\\
&  m_{2,4}^{(H)}=\frac{1}{4}\,,\quad m_{1,3,5}^{(H)}=m_{2,4}^{(I)}%
=m_{1,2,3,5}^{(J)}=\frac{1}{2}\,,\nonumber\\
&  m_{1,3}^{(I)}=m_{1,2,3}^{(K)}=m_{5}^{(L)}=1\,. \label{Gew-line}%
\end{align}
Using these weights as well as the symmetries of the diagrams, we obtain form
Eq.~(\ref{AvMom}) the following averaged moments of the weights:
\begin{align}
L_{av}^{(\alpha)}(A-2B,\{s_{i}\})  &  =-\Big(1-\frac{1}{2^{\alpha}%
}\Big)\bigl(s_{1}+s_{2}\bigr)\,,\nonumber\\
L_{av}^{(\alpha)}(C-4D-E+2F+4G,\{s_{i}\})  &  =\Big(1-\frac{1}{2^{\alpha}%
}\Big)\bigl(s_{1}+s_{2}+s_{3}+s_{4}\bigr)+\Big(2-\frac{3}{2^{\alpha}%
}\Big)s_{5}\,,\nonumber\\
L_{av}^{(\alpha)}(H-I-2J+2K+L,\{s_{i}\})  &  =\Big(1-\frac{1}{2^{\alpha}%
}\Big)\bigl(s_{1}+s_{3}+s_{5}\bigr)+\Big(1-\frac{1}{2^{\alpha}}\Big)^{2}%
\bigl(s_{2}+s_{4}\bigr)\,.
\end{align}

In the following, we replace the control parameter $w$ by $v_{\alpha}$ to
emphasize the fact that the multifractal index $\alpha$ is kept general in our
calculation. Using the substitutions $s_{1}=xt$, $s_{2}=(1-x)t$, the of the
$1$-loop self-energy that is linear in $v_{\alpha}$ becomes
\begin{align}
I_{0,v}  &  =-i\lambda v_{\alpha}\frac{g^{2}}{2}\int_{0}^{\infty}ds_{1}%
ds_{2}\frac{\exp\bigl[-(s_{1}+s_{2})\tau\bigr]}{(s_{1}+s_{2})^{d/2}}%
L_{av}^{(\alpha)}(A-2B,\{s_{i}\})\nonumber\\
&  =i\lambda v_{\alpha}\frac{g^{2}}{2(4\pi)^{d/2}}\int_{0}^{\infty
}dt\,t^{2-d/2}\exp(-\tau t)\,\Big(1-\frac{1}{2^{\alpha}}\Big)\nonumber\\
&  =i\lambda v_{\alpha}\frac{G_{\varepsilon}g^{2}}{\varepsilon}\tau
^{-\varepsilon/2}\Big(1-\frac{1}{2^{\alpha}}\Big)\,.
\end{align}
Using the renormalization scheme~(\ref{RG-Schema}) with the $w$ renormalization
replaced by
\begin{align}
 v_{\alpha}\rightarrow\mathring{v}_{\alpha}%
=Z^{-1}Z_{\alpha}v_{\alpha}\,,
\end{align}
we get
\begin{align}
\left.  \Gamma_{2}\right\vert _{v}^{(1l)}  &  =Z\left.  \mathring{\Gamma}%
_{2}\right\vert _{v}^{(1l)}=i\lambda v_{\alpha}Z_{\alpha}%
\Big\{1-Z^{-3+\varepsilon/2}Z_{\tau}^{-\varepsilon/2}Z_{u}\Big(1-\frac
{1}{2^{\alpha}}\Big)\frac{u}{\varepsilon}\Big(\frac{\mu^{2}}{\tau
}\Big)^{\varepsilon/2}\Big\}\nonumber\\
&  =i\lambda v_{\alpha}Z_{\alpha}\Big\{1-\Big[1+\Big(\frac{7}{2\varepsilon
}-\frac{5}{12}\Big)u\Big]\Big(1-\frac{1}{2^{\alpha}}\Big)\frac{u}{\varepsilon
}\Big(\frac{\mu^{2}}{\tau}\Big)^{\varepsilon/2}+O(u^{3})\Big\}\,.
\end{align}
for the $v_{\alpha}$-part of the renormalized vertex-function to $1$ -loop order. It follows %
\begin{equation}
Z_{\alpha}=1+\Big(1-\frac{1}{2^{\alpha}}\Big)\frac{u}{\varepsilon}+O(u^{2})
\end{equation}
for the renormalization factors of the $v_{\alpha}$.

Now, we turn to $2$-loop order. First, we consider the diagrams C, D, E, F, G.
These diagrams lead to the integral%
\begin{equation}
I_{1,v}=-i\lambda v_{\alpha}\frac{g^{4}}{2(4\pi)^{d}}\int\prod_{i=1}^{5}%
ds_{i}\,\frac{L_{av}^{(\alpha)}(C-4D-E+2F+4G,\{s_{i}\})\exp\bigl(-\tau\sum
_{i}s_{i}\bigr)}{\bigl[(s_{1}+s_{2})(s_{3}+s_{4})+s_{5}(s_{1}+s_{2}%
+s_{3}+s_{4})\bigr]^{d/2}}\,.
\end{equation}
Switching to the integration variables defined in Eq.~(\ref{varChange1}) and
integrating over $x$ and $y$, the integrals $I_{1,v}$ can be expressed once
again in terms of the mother-integral $M$. Taking the appropriate derivatives
thereof, we obtain
\begin{align*}
I_{1,v} &  =-i\lambda v_{\alpha}\frac{g^{4}}{2(4\pi)^{d}}\int\prod_{i=1}%
^{3}dt_{i}\,\frac{t_{1}t_{2}\exp\bigl[-\tau(t_{1}+t_{2}+t_{3})\bigr]}%
{\bigl(t_{1}t_{2}+t_{2}t_{3}+t_{3}t_{1}\bigr)^{d/2}}\\
&  \qquad\qquad\times\Big[2\Big(1-\frac{1}{2^{\alpha}}\Big)(t_{1}%
+t_{2})+\Big(2-\frac{3}{2^{\alpha}}\Big)t_{3}\Big]\\
&  =-i\lambda v_{\alpha}\frac{G_{\varepsilon}^{2}g^{4}}{\varepsilon}%
\tau^{-\varepsilon}\Big\{\Big(1-\frac{1}{2^{\alpha}}\Big)\frac{2}{\varepsilon
}+\Big(1-\frac{5}{2^{\alpha+2}}\Big)\Big\}\,.
\end{align*}
Proceeding similarly, we get for the second group of diagrams H, I, J, K, L%
\begin{equation}
I_{2,v}=-i\lambda v_{\alpha}\frac{g^{4}}{2(4\pi)^{d}}\int\prod_{i=1}^{5}%
ds_{i}\,\frac{L_{av}^{(\alpha)}(H-I-2J+2K+L,\{s_{i}\})\exp\bigl(-\tau\sum
_{i}s_{i}\bigr)}{\bigl[(s_{1}+s_{3}+s_{5})(s_{2}+s_{4})+s_{2}s_{4}%
\bigr]^{d/2}}\,.
\end{equation}
Using the integration variables defined in Eq.~(\ref{varChange2}), we obtain
\begin{align}
I_{2,v} &  =-i\lambda v_{\alpha}\frac{g^{4}}{4(4\pi)^{d}}\int\prod_{i=1}%
^{3}dt_{i}\,\frac{t_{1}^{2}\exp\bigl(-\tau(t_{1}+t_{2}+t_{3})\bigr)}%
{\bigl[t_{1}t_{2}+t_{2}t_{3}+t_{3}t_{1}\bigr]^{d/2}}\nonumber\\
&  \qquad\qquad\times\Big[\Big(1-\frac{1}{2^{\alpha}}\Big)t_{1}+\Big(1-\frac
{1}{2^{\alpha}}\Big)^{2}(t_{2}+t_{3})\Big]\nonumber\\
&  =-i\lambda v_{\alpha}\frac{G_{\varepsilon}^{2}g^{4}}{4\varepsilon}%
\tau^{-\varepsilon}\Big(1-\frac{1}{2^{\alpha}}\Big)\Big\{\Big(1-\frac
{1}{2^{\alpha-1}}\Big)\frac{1}{\varepsilon}-\Big(\frac{7}{4}+\frac
{1}{2^{\alpha+1}}\Big)\Big\}\,.
\end{align}
Collecting the $1$- and $2$-loop contributions, we find%
\begin{align}
\left.  \Gamma_{2}\right\vert _{v}^{(2l)} &  =Z\left.  \mathring{\Gamma}%
_{2}\right\vert _{v}^{(2l)}=i\lambda v_{\alpha}\bigg\{Z_{\alpha}%
-\Big(1-\frac{1}{2^{\alpha}}\Big)\Big[1+\Big(\frac{9}{2}-\frac{1}{2^{\alpha}%
}-\frac{5\varepsilon}{12}\Big)\frac{u}{\varepsilon}\Big]\frac{u}{\varepsilon
}\Big(\frac{\mu^{2}}{\tau}\Big)^{\varepsilon/2}\nonumber\\
&  +\Big[\Big(1-\frac{1}{2^{\alpha}}\Big)\Big(\frac{9}{2}-\frac{1}{2^{\alpha}%
}\Big)+\varepsilon\Big(\frac{9}{8}-\frac{15}{2^{\alpha+3}}+\frac{1}%
{4^{\alpha+1}}\Big)\Big]\frac{u^{2}}{2\varepsilon^{2}}\Big(\frac{\mu^{2}}%
{\tau}\Big)^{\varepsilon}\biggr\}+O(u^{3})\biggr\}\nonumber\\
&  =i\lambda v_{\alpha}\bigg\{1-\Big(1-\frac{1}{2^{\alpha}}\Big)\frac
{u}{\varepsilon}-\Big[\Big(\frac{9}{\varepsilon}-\frac{47}{12}\Big)-\Big(\frac
{11}{\varepsilon}-\frac{65}{12}\Big)\frac{1}{2^{\alpha}}+\Big(\frac
{2}{\varepsilon}-\frac{1}{2}\Big)\frac{1}{4^{\alpha}}\Big]\frac{u^{2}%
}{4\varepsilon}+O(u^{3})\biggr\}
\end{align}
for the $v_\alpha$-part of the renormalized self-energy to order $u^{2}$. It is
free of non-primitive divergencies as it should, and the $\varepsilon$-poles
are cancelled by the renormalization factors stated in
Eq.~(\ref{renFaktorZalpha}).
\end{widetext}

%\section{References}


\begin{thebibliography}{99}
\bibitem{footnote1}This step also helps to avoid factors like $\bigl(\delta(0)\bigr)^{D}$.
When interpreted properly, i.e., when  $\delta(0)$ is interpreted as the limit
of a finite quantity and the replica limit $D\rightarrow0$ is taken at first
these factors reduce to unity and are hence harmless. Nevertheless, avoiding
these factors helps to keep calculations clutter free.
\bibitem {Bemerkung}Note that in the figure shown in reference
\cite{FeBlFoHo04} a factor $1/36$ of the second order term is missed.
\end{thebibliography}
\end{document}